# The impact of population demography and selection on the genetic architecture of complex traits

Kirk E. Lohmueller[a]

[a]Department of Ecology and Evolutionary Biology, Interdepartmental Program in Bioinformatics,

University of California, Los Angeles, CA, 90095

**Address correspondence to:**

Kirk E. Lohmueller

Department of Ecology and Evolutionary Biology

University of California, Los Angeles

621 Charles E. Young Drive South

Los Angeles, CA 90095-1606

Office Phone: (310)-825-7636

Fax: (310)-206-0484

E-mail: klohmueller@ucla.edu

**ABSTRACT**

Population genetic studies have found evidence for dramatic population growth in recent human history. It is unclear how this recent population growth, combined with the effects of negative natural selection, has affected patterns of deleterious variation, as well as the number, frequency, and effect sizes of mutations that contribute risk to complex traits. Because researchers are performing exome sequencing studies aimed at uncovering the role of low-frequency variants in the risk of complex traits, this topic is of critical importance. Here I use simulations under population genetic models where a proportion of the heritability of the trait is accounted for by mutations in a subset of the exome. I show that recent population growth increases the proportion of nonsynonymous variants segregating in the population, but does not affect the genetic load relative to a population that did not expand. Under a model where a mutation's effect on a trait is correlated with its effect on fitness, rare variants explain a greater portion of the additive genetic variance of the trait in a population that has recently expanded than in a population that did not recently expand. Further, when using a single-marker test, for a given false-positive rate and sample size, recent population growth decreases the expected number of significant associations with the trait relative to the number detected in a population that did not expand. However, in a model where there is no correlation between a mutation's effect on fitness and the effect on the trait, common variants account for much of the additive genetic variance, regardless of demography. Moreover, here demography does not affect the number of significant associations detected. These findings suggest recent population history may be an important factor influencing the power of association tests and in accounting for the missing heritability of certain complex traits.

**AUTHOR SUMMARY**

Many human populations have dramatically expanded over the last several thousand years. I use population genetic models to investigate how recent population expansions affect patterns of mutations that reduce reproductive fitness and contribute to the genetic basis of complex traits (including common disease). I show that recent population growth increases the proportion of mutations found in the population that reduce fitness. When mutations that have the greatest effect on reproductive fitness also have the greatest effect on a complex trait, more of the heritability of the trait is due to mutations at very low-frequency in populations that have recently expanded, as compared to populations that have not. Also, under this model, for a given sample size and false-positive rate, fewer variants show statistically significant associations with the trait in the population that has expanded than in one that has not. Both of these findings suggest that recent population growth may make it more difficult to fully elucidate the genetic basis of complex traits that are directly or indirectly correlated with fitness.

**INTRODUCTION**

Genome-wide association studies (GWAS) have successfully detected associations between hundreds of common single nucleotide polymorphisms (SNPs) and complex traits in humans [1, 2]. While this catalog of genes has revealed important biological insights, for most traits, the discovered associations can only account for a small fraction of the heritability for these traits measured from family-based studies [3]. This difference in the heritability observed in familial studies and the heritability explained by associated SNPs has been termed "missing heritability", and there is tremendous interest in the human genetics community to find it [3-5].

One possibility that has received particular attention is that the missing heritability lies in rare variants that have large effect sizes [3]. Because a risk variant is rare in the population, an association between the variant and the phenotypes of interest may not have been detected using traditional GWAS. Instead, at present, such variants must be assayed through direct sequencing. Due to technological advances (e.g. next-generation sequencing) [6, 7], combined with newer analytical methods designed for analyzing full sequence data [8-11], exome and full genome-sequencing studies are now being implemented in human genetics. The progression to sequencing data has already proved fruitful for the identification of causal mutations for several Mendelian diseases [6, 12-16]. Full sequence data [17, 18] is starting to reveal a richer picture of low-frequency genetic variation (minor allele frequency <0.5%), which may, in turn, increase the community's ability to implicate rare variants in risk of complex disease [4, 19-25]. Further, such studies should allow researchers to empirically determine the extent to which rare variants account for the missing heritability of complex traits [26-30]. However, before these new technological and methodological advances can reach their full potential, a more thorough understanding of low-frequency genetic variation in multiple human populations is essential.

To learn about patterns of rare genetic variation, several studies have sequenced hundreds of genes or complete exomes in thousands of individuals [31-34]. These studies have made two important discoveries. First, they have found a larger number of rare variants than was expected under previous models of human population history. It has been argued that this excess of rare variants can be explained by the recent explosion in human population size [31, 32, 35-37]. Second, these studies have documented a plethora of rare nonsynonymous SNPs that are likely evolutionarily deleterious and may be of medical relevance.

Comparatively less work has been done, however, to examine the implications that recent population history has had on the architecture of complex traits (but see the recent manuscript by Simons et al. [38]). It is unclear whether population history, and recent population growth in particular, affects the number, frequency, and effect sizes of mutations that contribute risk to complex traits. Addressing this question is critical for finding the "missing heritability" in different populations, as well performing the most powerful association studies to implicate specific variants in disease risk. Over a decade ago, it was recognized that the power to associate common variants with complex disease varied across populations [39-41]. This was largely due to asymmetry in the extent of linkage disequilibrium (LD) across populations as a result of differences in demographic history [42-44]. While the issue of LD is less relevant when considering rare variants, the topic of population choice for association studies has received substantially less attention when considering rare variants, despite its potential importance.

Here I use population genetic models to investigate the effect of recent population growth on patterns of deleterious genetic variation, the architecture of complex traits, and the ability to associate causal variants with the trait in models where a proportion of the trait's heritability is accounted for by mutations in a subset of the exome. Specifically, I show that recent population growth increases the input of deleterious mutations into the population, directly causing a proportional excess of deleterious genetic variation segregating in the population. Second, if a mutation's effect on reproductive fitness is correlated with its effect on a complex trait (such as a disease), I show that recent population growth increases the amount of the additive genetic variance of the trait that is accounted for by low-frequency variants relative to that in a population that did not expand recently. Further, I demonstrate that recent population growth leads to an increase in the number of alleles that contribute to the trait relative to what is expected in a population that did not recently expand. Finally, recent population growth decreases the number of SNPs that are significantly associated with the trait, relative to the number detected in a population that did not recently expand. This work indicates that in certain circumstances, recent population history will play an important role in determining the genetic architecture of complex traits in a particular population under study. As such, recent population history is a factor that should be considered when designing and interpreting re-sequencing studies for complex traits.

**METHODS**

*Models of population history*

I explore several of models of population history (**Fig. 1**). Because many studies have inferred a population bottleneck in non-African human populations associated with the Out-of-Africa migration process [45-50], the first model includes a brief, but severe, reduction in population size (**Fig. 1A**). After the bottleneck, the population returns to the same size as the ancestral population. This model is referred to as "BN" throughout the rest of paper. The second model of population history also includes the same Out-of-Africa population bottleneck, but now includes an instantaneous, 100-fold population expansion in the last 80 generations, or the last 2000 years, assuming 25 years/generation (**Fig. 1B**) [51]. This recent explosion in effective population size is meant to approximate the expansion detected in the archeological and historical records as well as in studies of genetic variation [31, 35, 36, 52]. This model is referred to as "BN+growth" throughout the paper. Finally, for comparison purposes, I also investigate a model where a population experienced an ancient 2-fold expansion (**Fig. 1C**). Such a model is meant to reflect the history of African populations [46, 47, 53] and is referred to as "Old growth" in the paper.

*Forward simulations*

All results were obtained using the forward-in-time population genetic program described in Lohmueller et al. [54], with minor modifications. Briefly, the program assumes a Wright-Fisher model of population history. Each generation, alleles change frequency stochastically based on binomial sampling and deterministically based on the standard selection equations. Also in each generation, a Poisson distributed number of new mutations enter the population at rate $\theta/2$, where $\theta = 4N_i\mu$. Here $N_i$ is the population size in the *i*th epoch of population history (see below) and $\mu$ is the per-chromosome per generation mutation rate across all the coding sequence that was simulated. I set $\mu = 0.056$ for synonymous sites. For nonsynonymous sites, $\mu$ is 2.5 times higher, because of the larger number of sites that, when mutated, give rise to a nonsynonymous mutation. Selection coefficients for new mutations are drawn from a gamma distribution with the parameters as inferred in Boyko et al. [46]. As done in the Poisson Random Field framework, all mutations are assumed to be independent of each other [55].

Step-wise population size changes are included in the model by changing the population size ($N$) at particular time points. Size changes affect the number of mutations that enter the population during each individual epoch and the magnitude of genetic drift.

For computational efficiency, I divided the population size by 2 and rescaled all times to be two fold smaller than under the specified model. However, I keep the population scaled mutation rate ($\theta$), and the population scaled selection coefficient ($\gamma = 2Ns$), equal to the same values as for the larger population. This rescaling is possible because, in the diffusion limit, patterns of genetic variation only depend on the scaled parameters. Such a rescaling is customary in other forward simulation programs [56, 57]. Samples of 1000 chromosomes are taken from the population at different time points to calculate how diversity statistics change over time.

*Models of disease*

To evaluate the effect of recent demographic history on the architecture of complex traits, I simulate individuals who have a quantitative trait under various demographic scenarios. I assume that deleterious (nonsynonymous) mutations in a given mutational target account for some of the heritability of the trait. Such a model is implicitly assumed in exome re-sequencing studies used to implicate rare variants in disease risk. This quantitative trait could represent a trait that is measured on a quantitative scale (e.g. lipid levels) or represent the underlying risk to a dichotomous phenotype (e.g. diabetes). Below I provide a description of the model and parameters.

I investigate models where mutations at a subset of nonsynonymous sites can account for 5%, 10% or, 30% of the variance of the phenotype (i.e. the heritability accounted for by these variants is 5%, 10% or 30%). Many complex traits have heritabilties around 30% [79-82]. Thus, the models considered here assume that some fraction of this heritability is accounted for by variants within the mutational target (i.e. a portion of the exome) while the rest is accounted for by variants not modeled here (i.e. noncoding portions of the genome). The mutational target size, *M*, is the number of nonsynonymous sites in the genome that, if mutated, would generate a variant that affects the phenotype. Assuming a mutation rate of 1 x $10^{-8}$ per site per generation, I investigate mutational target sizes of 70 kb and 140 kb. To gain a sense of how these sites could be partitioned into genes, the median length of the coding region of human genes 1335 bp [58]. Thus, a random gene would have roughly 934 nonsynonymous sites (assuming 70% of the coding sites are nonsynonymous). If all nonsynonymous sites within the gene would, if mutated, produce a causal variant, then the mutational target size of 70kb would correspond to 75 distinct genes accounting for the specified heritability, and the target size of 140kb would correspond to 150 distinct causal genes accounting for the heritability. If only half the nonsynonymous sites could be mutated to causal variants, then the number of genes would increase by a factor of 2. In practice, this model is implemented by taking a subset of the nonsynonymous SNPs simulated as described above and then assigning them an effect on the trait.

To assign an effect on the trait to a given causal SNP, I follow the model described by Eyre-Walker [59], with a modification described below. Essentially, the $i^{th}$ SNP's effect on a trait, $\alpha_i$, is given by

$$\alpha_i = \delta s_i^\tau (1 + \varepsilon_i) C ,$$

where $\delta = 1$, $s_i$ is the selective disadvantage for the $i^{th}$ SNP, $\tau$ is the relationship between the SNP's effect on fitness and the trait. A value of $\tau = 1.0$ indicates a linear relationship, where the mutations that are most deleterious will also have the biggest effects on the trait. A value of $\tau = 0.0$ indicates that a mutation's effect on fitness is independent of its effect on the trait. I set $\tau = $ 0.5 and $\tau = 0.0$, to model a situation where there is some correlation between fitness and the trait, and another situation where the trait is independent of fitness. Next, $\varepsilon_i$ for the $i^{th}$ SNP is drawn from a normal distribution with mean 0 and a standard deviation of 0.5. I did not vary this standard deviation, because Eyre-Walker showed that varying this parameter had little effect on the overall results [59]. $C$ is a normalizing constant for the SNP effect sizes so that

$$V_A = \sum_{i \text{ SNPs}} 2p_i(1-p_i)\alpha_i^2 \approx h_C^2,$$

where $h_C^2 \in \{0.05, 0.1, 0.3\}$. Essentially, $C$ is a scaling constant for the SNP effect sizes so that the desired heritability is achieved under each combination of parameters $h_C^2$, $\tau$, and $M$. Importantly, I find the average value of $C$ across all simulation replicates in the standard neutral model, and then use this value of $C$ for simulations under the other demographic models. As such, a SNP with a given effect on the trait under one demographic scenario will have the same effect on the trait under a different demographic scenario. This framework has the desirable property that a SNP's effect on a trait in a particular individual is biologically determined and is not directly affected by the demography of the population. Additionally, when setting up the simulations in this manner, the actual $h^2$ in a given simulation replicate is the outcome of a stochastic process, rather than set to a specific value. Nevertheless, in practice, there was little variation in $h^2$ across different demographic scenarios (**Fig. S3**). Incidentally, different values of $C$ are found when using different values of $h_C^2$, $\tau$, and $M$ (**Table S1**). This is reasonable because these models are biologically very different from each other.

I then assign trait values ($Y_j$) to each simulated individual. This is done using an additive model,

$$Y_j = \sum_{i \text{ SNPs}} z_{ij}\alpha_i + \varepsilon_j,$$

where the summation is over all $i$ causal variants, $z_{ij}$ is the number of copies of the risk allele ($z_{i,j} \in \{0,1,2\}$) carried at the $i^{th}$ SNP by the $j^{th}$ individual, $\alpha_i$ is the effect of the $i^{th}$ SNP, and $\varepsilon_j$ is the environmental variance, which is drawn from a normal distribution with mean 0 and variance $1 - h_C^2$ (see [60-62]). For some analyses, I translate these quantitative traits into dichotomous diseases. To do this, I assume the liability threshold model for complex traits, where there is an underlying continuous distribution of risk in the population, and where cases are those individuals whose risk ($Y_j$) falls above a discrete threshold ($L$) [60, 63, 64]. $L$, or the liability threshold, was set in each simulation replicate by transforming the phenotypes ($Y_j$s) to follow the standard normal distribution and then picking the threshold such that 40% of the individuals had liabilities greater than $L$. In this model, the disease had a prevalence of 40%. 1000 case individuals were randomly sampled from the individuals in this upper tail of the distribution. 1000 controls were selected from the lower 60% of the distribution. Single marker association tests were then performed for each SNP using Fisher's exact test. A test was considered significant if its $P$-value was $<1 \times 10^{-5}$, unless otherwise stated.

**RESULTS**

*Recent growth and deleterious variation*

First I assess the effect that different population histories (BN, BN+growth, and Old Growth) have on neutral and deleterious genetic variation. **Fig. 2A** and **Fig. 2B** show how the number of synonymous and nonsynonymous, respectively, SNPs segregating in a sample of 1,000 chromosomes changes over time as the simulated populations change in size. The population bottleneck 2000 generations ago resulted in a decrease in the number of SNPs segregating in the BN and BN+growth populations (orange and green lines in **Fig. 2A** and **Fig. 2B**). When the populations recovered from the bottleneck and increased in size, the number of SNPs in the population also increased. This increase in the number of SNPs after the recovery from the bottleneck is due to two factors. First, the larger population size allows more new mutations to enter the population. Second, genetic drift has a weaker effect when the population size is large. As such, more SNPs are maintained in the population. The recent explosion in population size (dashed green lines in **Fig. 2A** and **Fig. 2B**; BN+growth) rapidly results in a substantial increase in the number of both synonymous and nonsynonymous SNPs segregating in the population. This is due to the extreme increase in the population mutation rate (typically referred to as $\theta$) due to the larger population size. Ancient population growth also resulted in an increase in the number of synonymous and nonsynonymous SNPs segregating in the population, via the same mechanisms (purple line in **Fig. 2A** and **Fig 2B**; Old growth).

**Fig. 2C** shows how the proportion of nonsynonymous SNPs segregating in the population changes over time. When populations BN and BN+growth decreased in size during the bottleneck, the proportion of nonsynonymous SNPs in the population also dropped (orange and green lines in **Fig. 2C**). The reason for this is that, when the population size decreases, rare variants are preferentially lost over common variants. More nonsynonymous than synonymous SNPs are rare, and, as such, the crash in population size results in the loss of more nonsynonymous SNPs than synonymous SNPs. After the population recovers from the bottleneck, the proportion of nonsynonymous SNPs found in the population increases (**Fig. 2C**). The reason for this increase is that, due to the increase in population size, many new mutations enter the population after the recovery from the bottleneck. Most of these new mutations are nonsynonymous, due to there being more possible nonsynonymous changes than synonymous changes in coding regions. In fact, the proportion of nonsynonymous SNPs segregating in the population immediately after the recovery of the bottleneck is actually higher than that in the ancestral population (**Fig. S1**). The very recent increase in size in the BN+growth population also results in an increase in the proportion of nonsynonymous SNPs (green line in **Fig. 2C**). 54.8% of the SNPs in BN+growth are nonsynonymous (green line in **Fig. 2C**), compared to 52.8% in the BN population (orange line in **Fig. 2C**). It will take approximately 4*Ne* (where *Ne* is the current effective population size) generations for the proportion of deleterious SNPs to reach the equilibrium value for the larger population size (**Text S1**). The population that underwent an ancient expansion (dotted purple line in **Fig. 2C**) also experienced an initial increase in the proportion of nonsynonymous SNPs segregating immediately after the expansion. However, because the expansion occurred long ago, selection has had sufficient time to remove many of the nonsynonymous SNPs and bring the proportion of nonsynonymous SNPs in the population below the value seen in the bottlenecked populations, consistent with previous simulations and empirical observations [54]. These results suggest that recent extreme changes in demographic history can have an impact on patterns of deleterious mutations that are segregating in the population. This pattern also holds with other magnitudes of population growth (**Text S1**).

Next I examine the average fitness effects of nonsynonymous SNPs segregating in a sample of 1,000 chromosomes at different time points in the simulations (**Fig. 2D**). During the bottleneck, the average segregating SNP in the BN and BN+growth populations becomes less deleterious than in the ancestral population (orange and green lines in **Fig. 2D**). This is due to many rare, deleterious SNPs being eliminated from the population as well as fewer new deleterious SNPs entering the population when it is small in size. After the population recovers from the bottleneck, the average segregating SNP became more deleterious. In the first few generations after the recovery, the average SNP is even more deleterious than in the ancestral population. This is due to the increase in the input of deleterious mutations immediately after the recovery from the bottleneck. After a few generations however, negative natural selection has eliminated many, though certainly not all, of these deleterious SNPs from the population. In fact, **Fig. 2D** shows that even in the present day, the average SNP is more deleterious than that in the ancestral population. This same effect applies even more strongly to the recent population growth within the last 80 generations. Immediately, after growth, the average SNP segregating in the BN+growth population was more strongly deleterious (**Fig. 2D**) than what is expected in a population that has not expanded. However, during the last 40 generations, selection has eliminated many of the most deleterious SNPs from the population. In the present day, the average SNP in the BN+growth population (green line) is slightly less deleterious the BN population (orange line, also see **Text S1**), consistent with the results of Gazave et al. [65]. This effect is less pronounced with decreasing amounts of population growth (**Text S1**).

The description of the average strength of selection on a SNP described above does not take into account the frequency of the deleterious SNP in the population. The genetic load, however, does by weighting the selection coefficient by the SNP's frequency [66]. Genetic load is the reduction in fitness of the population due to deleterious mutations [67]. I find that, unlike the average selection coefficient, the genetic load is not affected by the demographic history of the population (**Fig. S2**). Similar results have recently been reported by Simons et al. [38]. Thus, while the recent population growth increases the number of deleterious SNPs segregating in the population, this increase in load is offset by the fact that most of these new deleterious mutations are kept at very low frequency in the population. Put another way, while the load appears to be the same across demographic models, the way in which the populations arrive at that load differs across demographies. The BN+growth population contains many rare deleterious mutations. The BN populations contains fewer deleterious mutations, but those that are there tend to be at higher frequencies.

*Models of a complex trait are compatible with observed GWAS results*

One important question is to what extent recent population growth affects the architecture of complex traits and our ability to map the genes responsible for disease risk. To investigate this issue, I simulated quantitative phenotypes for individuals sampled from the simulations under the three different demographic models. I investigate two different models for the relationship between a mutation's effect on fitness (the selection coefficient), and its effect on the trait [59]. First, I assume that a mutation's effect on fitness is partially correlated with its effect on the trait ($\tau = 0.5$). Here, those mutations that are strongly deleterious have a greater effect on the trait. Second, I investigate a model where a SNP's effect on fitness is independent of its effect on the trait ($\tau = 0$). I also investigate different mutational target sizes ($M = 70$ kb and $M = 140$ kb) and the amount of the heritability accounted for by variants occurring within the mutational target ($h_C^2 \in \{0.3, 0.1, 0.05\}$; see Methods; see Discussion for further justification of these models).

While the model parameters were chosen to reflect what might be observed in exome resequencing data, there are few published well-powered exome sequencing studies for complex traits to which to test the validity of these parameter values. Thus, instead of comparing to sequencing data, I assess whether the models generated simulated datasets that were compatible with observations from GWAS studies. For simplicity, I make the assumption that the causal variants themselves have been directly assayed or have been imputed through LD with tag SNPs SNPs included in the GWAS. This of course is unlikely to be true in practice, particularly for rare variants [17, 18, 68]. Nevertheless, this comparison still serves as a useful benchmark to exclude models that are obviously not consistent with the observed GWAS data, with the caveat that some models that appear inconsistent with the GWAS data may actually fit better if LD and ascertainment bias were properly accounted for.

First, it has been consistently shown that the top SNPs identified through GWAS account for only a very limited amount of the phenotypic variance (often <10%) [3, 69]. I assess the amount of phenotypic variance ($V_P$) explained by the top 50 SNPs that account for the most variance (**Table S2**) within each simulation replicate. Models where $h_C^2 = 0.3$ and $M$=70 kb or 140 kb predict that the top 50 SNPs account for roughly 30% of the $V_P$, which appears to be too high to be compatible with most of the GWAS results, if one accepts the premise that GWAS would have detected the variants that explain such a large proportion of the phenotypic variance. However, a model with $h_C^2 = 0.05$ predicts that the top 50 SNPs will account for about 5% of the phenotypic variance. Such a model cannot be rejected from the currently available GWAS data.

Next, GWAS suggest that most risk loci for complex traits have very small effect sizes and are difficult to detect in samples of only 1000 cases and controls [69]. **Table S3** shows the expected number of GWAS hits ($P$<5 x 10$^{-8}$) expected in each simulation replicate in samples of 1000 cases and 100 controls for the different models of $h_C^2$, $M$, and $\tau$. Models where $h_C^2 = 0.3$ and $M$=70 kb or 140 kb predict that 1-4 significant GWAS hits should be observed. This is too many to be compatible with the observed data. Models with the mutational target accounting for less of the heritability ($h_C^2 = 0.1$ and $h_C^2 = 0.05$) predict <1 significant association. Thus, these models cannot be rejected based on GWAS data.

In summary, it is unclear how much of the heritability is accounted for by the exome and what the appropriate mutational target size for common diseases should be. Thus, I consider a variety of models examining different parts of this parameter space. Some of these models cannot be rejected on the basis of existing GWAS data. Other models may be more compatible with GWAS data if imperfect LD between causal variants and genotyped variants was properly taken into account. Overall, these models provide a framework consistent with existing empirical data with which to investigate the effect of recent population history and the genetic architecture of complex traits.

*Recent growth and the heritability of complex traits*

Using the models of demography, selection, and genetic architecture described above, I first examine the effect that population history has on the heritability of the trait. I find that population history has little effect on the heritability of the trait (**Fig. S3**), regardless of the values of $h_C^2$, $\tau$, and *M* used in the simulations. This is evidenced by the fact that in all three demographic scenarios investigated, the actual heritability estimated from each simulation replicate is close to $h_C^2$, the value set in a constant size population. To further investigate the effect of recent growth on heritability, I divide the causal variants segregating at the end of the simulation into three categories. The first category consists of those SNPs that arose either further back in time than, or during the population bottleneck ("Before bottleneck" in **Fig. 3**). These mutations occurred >1960 generations ago. The second category consists of SNPs that arose after the population had recovered from the bottleneck, but further back in time than the recent population growth ("After bottleneck" in **Fig. 3**). These mutations arose between 1960 and 80 generations ago. The final category consists of SNPs that arose within the last 80 generations ("After growth" in **Fig. 3**). In the BN+growth model, these are the mutations that arose after the population expansion. **Fig. 3A** shows that the average heritability accounted for by mutations that arose at these three different time points is similar in both the BN+growth population (green boxes), and the BN population (orange boxes). Interestingly, when a mutation's effect on fitness is correlated with its effect on the trait ($\tau = 0.5$), mutations that arose in the last 80 generations, as a class, account for the greatest amount of the heritability (**Fig. 3A**).

Next, I investigate whether other features of genetic variation that affect the heritability (e.g. number of SNPs, mean allele frequency, mean effect size) are affected by recent population history. I find that recent growth has had little effect on the number of mutations that arose prior to the population bottleneck and are still segregating in the sample, the mean allele frequency, and the effect sizes of such mutations (**Fig. 3B**, **Fig. 3C**, and **Fig. 3D**). However, there is a different pattern for mutations that arose after the bottleneck, but more than 80 generations ago (those in the "After bottleneck" category). Recent population growth increases the number of such mutations (roughly 2-fold) relative to that found in the population that did not expand (**Fig. 3B**). Further, these mutations tend to be at lower frequency in the BN+growth population compared to the BN population (**Fig. 3C**). The only difference between the two models of population history on variants that arose during this time period is that genetic drift is weaker in the BN+growth population, compared to the BN population. Thus, fewer weakly deleterious mutations are lost from the BN+growth population, generating the pattern seen in **Fig. 3B**. The mutations that are not lost from the population tended to be at lower frequency in the larger population because they also are less likely to drift to higher frequency in the expanded population as compared to the non-expanded populations. The mutations that arose within the last 80 generations also are affected by recent population history (those in the "After growth" category). As expected, recent population growth leads to a dramatic increase in the number of such SNPs (**Fig. 3B**). Further, the new mutations tend to be at lower frequency in the BN+growth population than in the BN population (**Fig. 3C**). More surprisingly, these SNPs tend to have weaker effect sizes on the trait in the BN+growth population than in the BN population (**Fig. 3D**). This observation can be explained by selection more effectively removing moderately and strongly deleterious mutations from the expanded population than from the non-expanded populations [65]. Thus, while recent population history affects the number of mutations, frequencies, and effect sizes of these mutations, it does so in such a way that the overall heritability of the trait appears unaffected by population history. **Fig. S4** shows similar plots for the model where a mutation's effect on the trait is not correlated with its effect on fitness ($\tau = 0$).

*Recent growth increases the contribution of rare variants to the additive genetic variance*

While population history does not affect the overall heritability of the trait, it can have a profound impact on the additive genetic variance ($V_A$), and consequently, the heritability, attributable to low-frequency vs. common variants (**Fig. 4**). When a mutation's effect on fitness is correlated with its effect on the trait ($\tau = 0.5$), more than 50% of the additive genetic variance in the trait is attributable to SNPs with frequency <0.5% in the population under all demographic scenarios, consistent with previous work showing of the importance of low-frequency variants [59, 70, 71]. Crucially, the amount of the variance attributable to rare variants (<0.1%) varies greatly due to demographic history. Roughly twice as much of the genetic variance in risk of the trait in the recently expanded population (BN+ growth; green line in **Fig. 4A**) is accounted for by SNPs with frequency <0.05% than in the population that underwent the bottleneck, but did not expand (BN; orange line in **Fig. 4A**). The population that underwent ancient growth falls intermediate to the other two cases (Old growth; purple line in **Fig. 4A**). Similar results hold for other heritabilities and mutational targets (**Fig. S5A**). The situation is dramatically different if a SNP's effect on the trait is uncorrelated with its effect on fitness ($\tau = 0$; **Fig. 4B**). Here little of the variance of the trait is accounted for by low-frequency variants, as seen by Eyre-Walker [59]. Additionally, under this model, demographic history does not make as substantial an impact on the amount of the additive genetic variance explained by SNPs at different frequencies, as suggested by Simons [38]. Again, similar results hold for other heritabilities considered and mutational target sizes (**Fig. S5B**). Thus, in some instances, recent population growth can result in a substantial increase in the amount of the genetic variance attributable to rare variants.

*Recent growth increases genetic heterogeneity of disease*

Population history also has a profound impact on the number of causal mutations in a sample of 1000 individuals who were selected from the upper $40^{th}$ percentile of the distribution of the quantitative trait (**Fig. 5**). These individuals can be thought of as cases. Here recent growth (BN+growth) is predicted to have resulted in a substantial increase in the number of causal mutations compared to a population that had not undergone such recent growth (BN; orange vs. green boxes in **Fig. 5A**). In fact, a sample of 1000 cases from the BN+growth population is predicted to have nearly twice as many distinct causal mutations as a sample from the BN population. An explanation for these patterns is that many new deleterious causal mutations have arisen after the population has expanded in size. Because they are new and rare, they are only found in a small number of individuals. As such, each individual has his/her own set of low-frequency risk mutations. When aggregating this number across hundreds of individuals, the total number of causal mutations in the sample from the BN+growth population is higher than in the BN population. Interestingly, the number of distinct causal mutations is actually higher in the BN+growth population than in the Old growth population (purple box in **Fig. 5A**).

A similar increase in the number of causal variants in the sample from the recently expanded population relative to a non-expanded population is seen even when a SNP's effect on the trait was uncorrelated with its effect on fitness ($\tau = 0$; **Fig. 5B**). This pattern is due to the fact that there is the same number of rare causal variants in the BN+growth population even when $\tau =$

0. However, when τ = 0, many of these rare causal mutations have smaller effect sizes, and consequently, do not account for much of the phenotypic variance of the trait.

To further explore this issue, I examine how much of the phenotypic variance ($V_P$) in the population can be accounted for by the SNPs that explain the most $V_P$ (**Fig. 6**). When τ = 0.5, the top SNPs that explain most of the variance account for less of it in the population that recently expanded (BN+growth) than in the population that did not (BN). For example, when $h_C^2 = 0.05$, the 25 SNPs that account for the most $V_P$ will account for 5% of the $V_P$ in the BN population (orange line in **Fig. 6A**). In contrast, for the BN+growth population (green line in **Fig. 6A**), the 25 SNPs that account for the most $V_P$ will only explain <3.5% of it. Put another way, the top 25 SNPs that explain the most variance account for >90% of the $V_A$ in the BN population, but <70% of the $V_A$ in the BN+growth population. These results suggest that, if mutational effects on disease are correlated with their effects on fitness, many of the additional rare causal variants found in a recently expanded population, may, in aggregate, explain a substantial proportion (say 20%) of the heritability of the trait.

If a mutation's effect on fitness is independent of its effect on disease (τ = 0), then the top SNPs that explain the most variance account for almost all of the $V_A$ (**Fig. 6B**). For example, in the model where $h_C^2 = 0.05$, the top 25 SNPs will account for nearly 5% of the $V_P$, regardless of the demographic history of the population. Put another way, here the top 25 SNPs account for the majority of the $V_A$, and this pattern is not affected by the demography of the population. This finding supports the previous statement that many of the extra causal mutations seen in **Fig. 5B** in the recently expanded population actually contribute very little to the overall $V_P$ of the trait. Similar results are found for other values of $h_C^2$ and mutational target sizes (**Fig. S6**).

*Effect of demography on the power of association tests*

Next, I investigate how different demographic histories affect the power to associate SNPs with a trait in a sample of 1000 cases and 1000 controls. Most power simulations for association tests examine the power to detect a given causal variant conditional on its allele frequency and/or effect size. Using this approach, I find that power to detect the SNPs that explain the greatest amount of $V_A$ is actually higher in the population that recently expanded (BN+growth) than in the population that only underwent a bottleneck (BN; **Text S2**). However, recent population growth has a more limited effect on the power to detect a given association when conditioning on the allele frequency or odds ratio of the causal SNP (**Text S2**).

The power analyses described above refer to the power to detect a given causal variant, conditional on various attributes of it. However, the number of causal variants, their frequencies, and their effect sizes are random quantities that are influenced by the evolutionary process experienced by the population under study. Thus, it is also useful to examine the expected number of causal SNPs with *P*-values less than the significance threshold across the different models of demographic history (**Fig. 7**). The expected number of causal SNPs detected in a study of a given sample size will account for both the power to detect a given variant as well as the number, frequency distribution, and effect size distribution of causal variants in the population. It is also directly answers the relevant question for researchers when planning and interpreting an association study: Under a given model of genetic architecture, with a given sample size, how many significant associations would I expect to detect?

Under the model where a mutation's effect on fitness is correlated with its effect on the trait ($\tau = 0.5$), $h_C^2 = 0.3$, and *M*=70kb, fewer causal mutations are detected in the populations that have undergone an ancient (Old growth; purple box in **Fig. 7A**) or recent (BN+growth; green box in **Fig. 7A**) expansion, relative to the population that only underwent a recent bottleneck (BN; orange box in **Fig. 7A**). Similar trends are seen for the other models of $h_C^2$ and $M$ (**Fig. S7**). However, when $h_C^2 = 0.05$, sample sizes of 1000 cases and controls are too small to detect almost any associations with $P<1 \times 10^{-5}$, regardless of the demography of the population. When using a less stringent significance threshold ($P<0.01$), $h_C^2 = 0.3$, and $M$=70kb, a median of 10 causal loci are associated with the trait in the BN population (**Fig. S8A**). However, a median of only 8 causal loci were detected in the BN+growth population. Again, similar trends are noted for the other models of $h_C^2$ and $M$ (**Fig. S8**). However, when $h_C^2 = 0.05$, a median of 2 causal SNPs were detected at $P<0.01$ for all three demographic models. This result is due to the very low power to detect an association for causal variants with very small effect sizes using samples of 1000 cases and 1000 controls, regardless of the demography of the population. Nevertheless, even here, a higher proportion of simulation replicates had detected at least 3 associations in the BN population (54%) than in the BN+growth population (41%). Taken together, this analysis suggests that recent population growth can result in a decrease in the expected number of associations to be detected in a given sample size. Thus, while recent growth may increase power to detect the SNP that explains the greatest amount of the variance, and have little effect on power to detect a given SNP conditional on its frequency or effect size, it enriches the frequency distribution for rare causal variants. The power to detect such variants using single-marker association tests is low, decreasing the expected number of significant association to be detected in the population that recently expanded.

However, demographic history has no clear effect on the number of causal loci detected with a given sample size when the mutation's effect on fitness is independent of its effect on the trait ($\tau = 0$; **Fig. 7B, Fig. S7**, and **Fig. S8**). For some models, the BN+growth population appears to have a higher number of significant associations than in the BN population (**Fig. S7D** and **Fig.**

**S7E**). However, this pattern is not consistently seen across models or significance thresholds. Similarly, when using a significance threshold of $P<0.01$, the Old growth population appears to show a greater number of significant association (**Fig. S8E-H**) than either of the other two models of population history. This pattern may be due to the slight, but noticeable, increase in the $h_C^2$ for the Old growth population (**Fig. S3E-H**).

Researchers have suggested that the amount of the additive genetic variance ($V_A$) explained by a set of SNPs increases when the stringency of the *P*-value threshold for including SNPs in the set is decreased [72-74]. **Fig. 8** shows the amount of additive genetic variance explained by SNPs having single-marker *P*-values less than the threshold specified on the *x*-axis for the model where $h_C^2 = 0.05$ and *M*=70kb. When a SNP's effect on the trait is correlated with its effect on fitness ($\tau = 0.5$), population history has an important effect on the amount of $V_A$ accounted for by SNPs with a given *P*-value (**Fig. 8A**). Specifically, recent population growth decreases the amount of $V_A$ accounted for by SNPs at all *P*-value thresholds, relative to what is seen in a population that has not expanded. For example, SNPs having $P<0.05$ account for about 30% of the $V_A$ in the BN population. SNPs with $P<0.05$ only account for 20% of $V_A$ in the BN+growth population. Including all SNPs detected in the case-control study only captures 70% of $V_A$ in the BN+growth population. The reason for this is that many of the rare variants that account for $V_A$ for the trait in the population were not present in the sample of 1000 cases and 1000 controls. When $\tau = 0$, population history has little affect on the amount of $V_A$ accounted for by SNPs with a given *P*-value (**Fig. 8B**). Including all SNPs present in the association study captures over 95% of the $V_A$. This finding is not surprising in light of the observation (**Fig. 4B**) that much of the $V_A$ is accounted for by common variants when $\tau = 0$, and such variants are likely to be present in the sample of 1000 cases and controls. Qualitatively similar trends are seen for other heritablities and mutational target sizes (**Fig. S9**).

**DISCUSSION**

I have shown that very recent population growth can have a profound impact on patterns of deleterious genetic variation and the genetic architecture of complex traits. Specifically, I show that recent population growth leads to an increase in the proportion of nonsynonymous SNPs relative to non-expanded populations. Further, this recent growth is predicted to have affected the genetic architecture of some complex traits. This result has implications for discovering the "missing heritability" in different human populations and detecting causal variants that may also affect reproductive fitness.

While it has been shown that differences in population history between European and African populations has affected the proportion of deleterious SNPs in the two populations [54], here I demonstrate that the influence of population history on deleterious mutations also applies on a much more recent timescale, and to populations that are much more similar to each other than Europeans and Africans.

While I have shown that demographic history greatly affects the proportion and frequencies of deleterious mutations segregating in the population, it is interesting that demography does not have a large effect on the overall genetic load of the population. Haldane has shown that the genetic load at equilibrium contributed by a particular mutation is independent of the strength of selection acting on the particular mutation and the frequency of that mutation [66]. Mutations of strong effect will be maintained by selection at lower frequency than mutations of weaker effect. Haldane suggests that these effects should cancel each other out. Haldane's result was derived for a simple model with a constant population size. It was unclear whether this result would hold when considering populations with bottlenecks and recent growth. Here I have shown that Haldane's result applies under certain complex demographic models. Further work is required to determine whether this trend holds in other species with demographic histories that depart even further from the standard neutral model, and whether this trend holds for models involving dominance.

I find that population history is predicted to have little effect on the overall amount of additive genetic variance for a trait seen in different populations. As such, assuming a common environmental variance across populations, the heritability of a trait is predicted to be similar across populations. This finding suggests that if differences in the heritability of a trait are detected across populations, these differences are more likely to be due to differing environmental effects, rather than due to different amounts of additive genetic variance. For example, it has been suggested that the heritability of height in a West African population is less than that typically estimated from European populations [75]. My results would argue that such a difference would be due to shifts in the environmental variance, rather than changes in amount of additive genetic variance as a result of differences in recent population history.

A major conclusion of this study is that recent population growth has a greater effect on the architecture of traits when a mutation's effect on fitness is correlated with its effect on the phenotype than when the mutation's effect on fitness is independent of its effect on the phenotype. The extent to which mutations that increase risk of disease are under purifying selection remains unclear. While it is intuitive that disease mutations should be under purifying selection, most common diseases have an onset after reproductive age, and as such, may not be correlated with fitness. However, it is possible that mutations increasing risk to late-onset disease may have pleiotropic effects and could affect traits related to reproduction [59, 76]. In fact, it has been suggested that pleiotropy is rather frequent for many common SNPs associated with disease [77], and may apply to rare variants as well. Additionally, genes and common variants associated with common diseases show signatures of purifying selection [78, 79], suggesting that disease variants may be under purifying selection. A third line of evidence comes from studies of model organisms. A mutagenesis study [80] found that *P*-element insertions that contributed the most to the variance in bristle number in *Drosophila* tended to reduce viability. A fourth line of evidence suggesting that a mutation's effect on complex disease may be correlated with fitness comes from an empirical analysis of GWAS data. Looking at over 350 susceptibility SNPs across eight categories of phenotypes, Park et al. [81] found that low-frequency SNPs tended to have larger effect sizes than more common SNPs (significantly so for type 1 diabetes, height, and LDL levels), even after correcting for the ascertainment bias resulting from the reduced power to detect associations with low-frequency SNPs of weak effect. A correlation between a mutation's effect on disease and its frequency is not expected under a model where a mutation's effect on disease is independent of its effect on fitness ($\tau = 0$; **Fig. S10B**). However, such a correlation is expected under models where a mutation's effect on fitness is correlated with its effect on the trait ($\tau = 0.5$; **Fig. S10A**). There is little direct evidence to indicate whether this result holds for low-frequency variants in coding regions. But, selection is likely to be stronger (per base pair) in coding regions than throughout noncoding regions of the genome [82], suggesting this result should hold for coding regions as well. Finally, the correlation between a mutation's effect on fitness and its effect on a trait is likely to depend on the particular trait involved. While further empirical and theoretical work is needed in this area, all of these lines of evidence suggest that, for some traits, it is plausible that a mutation's effect on fitness could indeed be correlated with its effect on disease.

Further rationale for considering models where a mutation's effect on fitness is correlated with its effect on the trait comes from exome sequencing studies themselves. A major assumption made in exome sequencing studies is that some of the missing heritability can be explained by rare variants of large effect that increase risk to disease [3, 4]. If there is no correlation between a mutation's effect on disease and its effect on fitness, then there is no reason for rare variants to have stronger effects on disease than more common variants. Under this model (where fitness effects are independent of trait effects), effect sizes would be randomly assigned to SNPs, regardless of their allele frequency. On the other hand, if a mutation's effect on fitness is correlated with its effect on disease, then the SNPs with the strongest effects on disease are likely to be the most deleterious ones. As such, they will also be the most rare in the population due to purifying selection. Because the exome sequencing paradigm essentially assumes that the effect of a coding region mutation on disease is correlated with its effect on fitness, it is important to investigate the proprieties of such a model under different population histories.

My models make several predictions that can be tested with empirical data. While these models were developed to apply to exome sequencing data, because the predictions were robust to the mutational target size and heritability accounted for by the mutations in the target region (**Fig. S3**, **Fig. S5**-**Fig. S9**), they should apply to GWAS data as well, especially if low-frequency variants are imputed from a reference panel, like the 1000 Genomes Project. First, the models predict that if a mutation's effect on fitness is correlated with its effect on the trait, common variants should account for more of the heritability in a population that did not expand than in one that had recently expanded. This prediction can be tested by analyzing GWAS data in expanded vs. non-expanded populations. Second, for a given sample size, if a mutation's effect on fitness is correlated with its effect on the trait, the models predict that fewer significant associations will be detected in the recently expanded population than in a population that has not expanded. This prediction can also be tested by comparing the number of significant associations detected in GWAS data from the expanded population vs. the non-expanded population. Third, the prediction that, if a mutations' effect on fitness is correlated with its effect on the trait, low-frequency variants should account for more of the heritability in the recently expanded population than in a non-expanded population can be tested directly once large-scale exome sequencing data in both expanded and non-expanded populations has been collected. Failing to find these patterns in GWAS and exome sequencing data would suggest that there is little correlation between a mutation's effect on fitness and its effect on the trait.

Several recent studies have used results from population genetic models to guide the design and interpretation of association studies of rare variants [83, 84]. My results are especially complementary to those Zuk et al. [84]. In particular, they argue that a population expansion does not increase the proportion of the disease due to new alleles. My finding of a similar contribution of young alleles to the heritability in expanded vs. non-expanded populations (**Fig. 3A**) supports their conclusion, despite different modeling assumptions in the two studies. Zuk et al. [84] also argue that the "role" of rare variants in disease is increased in an expanding population. Here I provide a more detailed analysis of this topic using an explicit quantitative genetics model and evaluate the conditions under which recent population growth accentuates the contribution of rare variants to the heritability of the trait.

One important limitation of the present study is that I evaluated the power of single-marker association tests, rather than gene-based association tests. It has been suggested that single-marker tests may be under-powered relative to gene-based tests for detecting associations with rare variants [37, 85]. I did not consider gene-based association tests in the present study because the present simulations assume that all SNPs are independent of each other. Thus, there is no way to simulate genes containing multiple SNPs having the appropriate LD structure in this framework. Future work could simulate larger genic regions using other approaches [56]. However, the analysis of the performance of single-marker association tests still provides important insights for understanding how demography affects mapping genes for complex traits. First, many association studies of rare variants include single-marker association tests, even if they also consider gene-based tests [23-25, 85]. Thus, our results are directly applicable to designing and interpreting such studies. Second, several gene-based association tests combine the single-marker association tests in various ways. Thus, the manner in which single-marker signals are affected by demography is relevant for such tests. Third, it is not clear that gene-based tests are always superior to single marker tests. A recent study [86] has suggested that gene-based tests based on single marker association statistics may be more powerful than other "burden tests". Further, performance of gene-based association tests is known to decrease if many non-causal SNPs are included [22, 84, 86]. Also, if causal variants are scattered across many distinct genes, then gene-based association tests may not provide an increase in power over single-marker tests [30]. As sample sizes continue to grow, it is likely that single marker tests will be more frequently used for sequencing-based association studies because they are simpler to implement and eliminate the need to decide how to combine variants within a gene. Thus, insights gained from the analysis of single-marker association tests should still be useful.

    Another possible limitation of this study is that my models do not allow some mutations affecting complex traits to be beneficial or under balancing selection. There is some evidence that loci associated with certain traits (obesity and type 2 diabetes, in particular [87, 88]) may

have been affected by positive selection. However, this pattern was found for common variants outside of coding regions. It is not clear how many nonsynonymous causal mutations would be expected to be under positive selection. Additionally, loci under positive or balancing selection should have already been detected in GWAS (because they would be common variants), and are not the type of loci researchers are aiming to discover through exome sequencing studies.

My results suggest that if a mutation's effect on fitness is correlated with its effect on the trait, recent population history can have an important effect on the ability to detect associations with causal variants. In the simulations, fewer causal SNPs were significantly associated with disease in the population that experienced recent growth as compared to the population that did not expand. This result would imply that, in order to detect the greatest number of causal loci for a given sample size, it would be better to focus on populations that experienced a bottleneck, but did not experience a recent population expansion. Currently, it is not clear which populations meet this criterion. Further sequencing of large samples of individuals will be required to determine which populations have not experienced recent population growth. For a variety of reasons, most GWAS have been done in large samples of cases and controls of European ancestry [89]. The same trend may hold for exome and genome re-sequencing studies. Recent genetic studies, as well as historical records indicate that many European populations are precisely those that experienced the type of extreme, recent population growth simulated in this study [36]. The simulations presented here suggest that focusing on such populations may not discover the largest number of causal variants for a given sample size of individuals sequenced.

An important goal in human genetics is to understand disease risk in populations throughout the globe. While focusing on populations that have not experienced ancient or recent growth may yield the largest list of putative causal loci, it is currently not clear whether such an approach will lead to increased understanding of the genetic basis of disease in all populations—not just the population under study. Analyses of common variants included in GWAS suggest that loci associated with traits in European populations also affect the traits in other non-European populations [90-93]. However, it is unclear whether this trend also holds for rare variants that have arisen after the populations have split. Thus, in order to understand the genetic basis of complex traits across the globe, it will be important to study populations that have recently expanded in addition to those more stable in size, recognizing that larger sample sizes in populations that have recently expanded will be necessary to achieve comparable power to that in non-expanded populations.

Finally, these results are directly relevant for finding the "missing heritability" in different populations. If a mutation's effect on disease is correlated with its effect on fitness, then more of the heritability will be explained by very rare variants in a population that experienced a recent expansion than in a population that did not recently expand. Additionally, the variants detected by single marker association tests explain less of the heritability in a recently expanded population than in a population that did not recently expand. Thus, while the overall heritability of a trait may not be variable across populations, our ability to discover the variants that account for it is likely to vary across populations due to differences in demographic history.

**ACKNOWLEDGEMENTS**

I thank Youna Hu, Montgomery Slatkin, Bogdan Passanuic, Alex Platt, Frank Albert, and members of the Lohmueller and Pasanuic labs for helpful discussions throughout this project and Clare Marsden, Frank Albert and the two reviewers for helpful comments on the manuscript.

# REFERENCES

1. Altshuler D, Daly MJ, Lander ES. (2008) Genetic mapping in human disease. Science 322: 881-888.
2. Stranger BE, Stahl EA, Raj T. (2011) Progress and promise of genome-wide association studies for human complex trait genetics. Genetics 187: 367-383.
3. Manolio TA, Collins FS, Cox NJ, Goldstein DB, Hindorff LA, et al. (2009) Finding the missing heritability of complex diseases. Nature 461: 747-753.
4. Gibson G. (2012) Rare and common variants: Twenty arguments. Nat Rev Genet 13: 135-145.
5. Eichler EE, Flint J, Gibson G, Kong A, Leal SM, et al. (2010) Missing heritability and strategies for finding the underlying causes of complex disease. Nat Rev Genet 11: 446-450.
6. Ng SB, Turner EH, Robertson PD, Flygare SD, Bigham AW, et al. (2009) Targeted capture and massively parallel sequencing of 12 human exomes. Nature 461: 272-276.
7. Shendure J, Ji H. (2008) Next-generation DNA sequencing. Nat Biotechnol 26: 1135-1145.
8. Wu MC, Lee S, Cai T, Li Y, Boehnke M, et al. (2011) Rare-variant association testing for sequencing data with the sequence kernel association test. Am J Hum Genet 89: 82-93.
9. Madsen BE, Browning SR. (2009) A groupwise association test for rare mutations using a weighted sum statistic. PLoS Genet 5: e1000384.
10. Li B, Leal SM. (2008) Methods for detecting associations with rare variants for common diseases: Application to analysis of sequence data. Am J Hum Genet 83: 311-321.
11. Li B, Leal SM. (2009) Discovery of rare variants via sequencing: Implications for the design of complex trait association studies. PLoS Genet 5: e1000481.
12. Bamshad MJ, Ng SB, Bigham AW, Tabor HK, Emond MJ, et al. (2011) Exome sequencing as a tool for mendelian disease gene discovery. Nat Rev Genet 12: 745-755.
13. Bamshad MJ, Shendure JA, Valle D, Hamosh A, Lupski JR, et al. (2012) The centers for Mendelian genomics: A new large-scale initiative to identify the genes underlying rare Mendelian conditions. Am J Med Genet A 158A: 1523-1525.
14. Ng SB, Bigham AW, Buckingham KJ, Hannibal MC, McMillin MJ, et al. (2010) Exome sequencing identifies *MLL2* mutations as a cause of Kabuki syndrome. Nat Genet 42: 790-793.
15. Ng SB, Buckingham KJ, Lee C, Bigham AW, Tabor HK, et al. (2010) Exome sequencing identifies the cause of a Mendelian disorder. Nat Genet 42: 30-35.
16. Ng SB, Nickerson DA, Bamshad MJ, Shendure J. (2010) Massively parallel sequencing and rare disease. Hum Mol Genet 19: R119-24.
17. 1000 Genomes Project Consortium, Durbin RM, Abecasis GR, Altshuler DL, Auton A, et al. (2010) A map of human genome variation from population-scale sequencing. Nature 467: 1061-1073.
18. 1000 Genomes Project Consortium, Abecasis GR, Auton A, Brooks LD, DePristo MA, et al. (2012) An integrated map of genetic variation from 1,092 human genomes. Nature 491: 56-65.
19. Cirulli ET, Goldstein DB. (2010) Uncovering the roles of rare variants in common disease through whole-genome sequencing. Nat Rev Genet 11: 415-425.
20. Gorlov IP, Gorlova OY, Sunyaev SR, Spitz MR, Amos CI. (2008) Shifting paradigm of association studies: Value of rare single-nucleotide polymorphisms. Am J Hum Genet 82: 100-112.
21. Schork NJ, Murray SS, Frazer KA, Topol EJ. (2009) Common vs. rare allele hypotheses for complex diseases. Curr Opin Genet Dev 19: 212-219.
22. Kiezun A, Garimella K, Do R, Stitziel NO, Neale BM, et al. (2012) Exome sequencing and the genetic basis of complex traits. Nat Genet 44: 623-630.
23. Helgason H, Sulem P, Duvvari MR, Luo H, Thorleifsson G, et al. (2013) A rare nonsynonymous sequence variant in *C3* is associated with high risk of age-related macular degeneration. Nat Genet 45: 1371-1374.

24. Seddon JM, Yu Y, Miller EC, Reynolds R, Tan PL, et al. (2013) Rare variants in *CFI*, *C3* and *C9* are associated with high risk of advanced age-related macular degeneration. Nat Genet 45: 1366-1370.
25. Zhan X, Larson DE, Wang C, Koboldt DC, Sergeev YV, et al. (2013) Identification of a rare coding variant in complement 3 associated with age-related macular degeneration. Nat Genet 45: 1375-1379.
26. Heinzen EL, Depondt C, Cavalleri GL, Ruzzo EK, Walley NM, et al. (2012) Exome sequencing followed by large-scale genotyping fails to identify single rare variants of large effect in idiopathic generalized epilepsy. Am J Hum Genet 91: 293-302.
27. Need AC, McEvoy JP, Gennarelli M, Heinzen EL, Ge D, et al. (2012) Exome sequencing followed by large-scale genotyping suggests a limited role for moderately rare risk factors of strong effect in schizophrenia. Am J Hum Genet 91: 303-312.
28. Hunt KA, Mistry V, Bockett NA, Ahmad T, Ban M, et al. (2013) Negligible impact of rare autoimmune-locus coding-region variants on missing heritability. Nature 498: 232-235.
29. Liu L, Sabo A, Neale BM, Nagaswamy U, Stevens C, et al. (2013) Analysis of rare, exonic variation amongst subjects with autism spectrum disorders and population controls. PLoS Genet 9: e1003443.
30. Lohmueller KE, Sparso T, Li Q, Andersson E, Korneliussen T, et al. (2013) Whole-exome sequencing of 2,000 Danish individuals and the role of rare coding variants in type 2 diabetes. Am J Hum Genet 93: 1072-1086.
31. Tennessen JA, Bigham AW, O'Connor TD, Fu W, Kenny EE, et al. (2012) Evolution and functional impact of rare coding variation from deep sequencing of human exomes. Science 337: 64-69.
32. Nelson MR, Wegmann D, Ehm MG, Kessner D, St Jean P, et al. (2012) An abundance of rare functional variants in 202 drug target genes sequenced in 14,002 people. Science 337: 100-104.
33. Fu W, O'Connor TD, Jun G, Kang HM, Abecasis G, et al. (2013) Analysis of 6,515 exomes reveals the recent origin of most human protein-coding variants. Nature 493: 216-220.
34. Marth GT, Yu F, Indap AR, Garimella K, Gravel S, et al. (2011) The functional spectrum of low-frequency coding variation. Genome Biol 12: R84-2011-12-9-r84.
35. Coventry A, Bull-Otterson LM, Liu X, Clark AG, Maxwell TJ, et al. (2010) Deep resequencing reveals excess rare recent variants consistent with explosive population growth. Nat Commun 1: 131.
36. Keinan A, Clark AG. (2012) Recent explosive human population growth has resulted in an excess of rare genetic variants. Science 336: 740-743.
37. Kryukov GV, Shpunt A, Stamatoyannopoulos JA, Sunyaev SR. (2009) Power of deep, all-exon resequencing for discovery of human trait genes. Proc Natl Acad Sci U S A 106: 3871-3876.
38. Simons YB, Turchin MC, Pritchard JK, Sella G. (2013) The deleterious mutation load is insensitive to recent population history. arXiv: 1305 2061v1 .
39. Wright AF, Carothers AD, Pirastu M. (1999) Population choice in mapping genes for complex diseases. Nat Genet 23: 397-404.
40. Pritchard JK, Przeworski M. (2001) Linkage disequilibrium in humans: Models and data. Am J Hum Genet 69: 1-14.
41. Peltonen L, Palotie A, Lange K. (2000) Use of population isolates for mapping complex traits. Nat Rev Genet 1: 182-190.
42. Service S, DeYoung J, Karayiorgou M, Roos JL, Pretorious H, et al. (2006) Magnitude and distribution of linkage disequilibrium in population isolates and implications for genome-wide association studies. Nat Genet 38: 556-560.
43. Ardlie KG, Kruglyak L, Seielstad M. (2002) Patterns of linkage disequilibrium in the human genome. Nat Rev Genet 3: 299-309.

44. Reich DE, Cargill M, Bolk S, Ireland J, Sabeti PC, et al. (2001) Linkage disequilibrium in the human genome. Nature 411: 199-204.
45. Akey JM, Eberle MA, Rieder MJ, Carlson CS, Shriver MD, et al. (2004) Population history and natural selection shape patterns of genetic variation in 132 genes. PLoS Biol 2: e286.
46. Boyko AR, Williamson SH, Indap AR, Degenhardt JD, Hernandez RD, et al. (2008) Assessing the evolutionary impact of amino acid mutations in the human genome. PLoS Genet 4: e1000083.
47. Gutenkunst RN, Hernandez RD, Williamson SH, Bustamante CD. (2009) Inferring the joint demographic history of multiple populations from multidimensional SNP frequency data. PLoS Genet 5: e1000695.
48. Keinan A, Mullikin JC, Patterson N, Reich D. (2007) Measurement of the human allele frequency spectrum demonstrates greater genetic drift in East Asians than in Europeans. Nat Genet 39: 1251-1255.
49. Lohmueller KE, Bustamante CD, Clark AG. (2009) Methods for human demographic inference using haplotype patterns from genomewide single-nucleotide polymorphism data. Genetics 182: 217-231.
50. Voight BF, Adams AM, Frisse LA, Qian Y, Hudson RR, et al. (2005) Interrogating multiple aspects of variation in a full resequencing data set to infer human population size changes. Proc Natl Acad Sci U S A 102: 18508-18513.
51. Tennessen JA, O'Connor TD, Bamshad MJ, Akey JM. (2011) The promise and limitations of population exomics for human evolution studies. Genome Biol 12: 127.
52. Gazave E, Ma L, Chang D, Coventry A, Gao F, et al. (2014) Neutral genomic regions refine models of recent rapid human population growth. Proc Natl Acad Sci U S A 111: 757-762.
53. Lohmueller KE, Bustamante CD, Clark AG. (2010) The effect of recent admixture on inference of ancient human population history. Genetics 185: 611-622.
54. Lohmueller KE, Indap AR, Schmidt S, Boyko AR, Hernandez RD, et al. (2008) Proportionally more deleterious genetic variation in European than in African populations. Nature 451: 994-997.
55. Sawyer SA, Hartl DL. (1992) Population genetics of polymorphism and divergence. Genetics 132: 1161-1176.
56. Hernandez RD. (2008) A flexible forward simulator for populations subject to selection and demography. Bioinformatics 24: 2786-2787.
57. Hoggart CJ, Chadeau-Hyam M, Clark TG, Lampariello R, Whittaker JC, et al. (2007) Sequence-level population simulations over large genomic regions. Genetics 177: 1725-1731.
58. Clamp M, Fry B, Kamal M, Xie X, Cuff J, et al. (2007) Distinguishing protein-coding and noncoding genes in the human genome. Proc Natl Acad Sci U S A 104: 19428-19433.
59. Eyre-Walker A. (2010) Evolution in health and medicine Sackler colloquium: Genetic architecture of a complex trait and its implications for fitness and genome-wide association studies. Proc Natl Acad Sci U S A 107 Suppl 1: 1752-1756.
60. Visscher PM, Hill WG, Wray NR. (2008) Heritability in the genomics era--concepts and misconceptions. Nat Rev Genet 9: 255-266.
61. Yang J, Lee SH, Goddard ME, Visscher PM. (2011) GCTA: A tool for genome-wide complex trait analysis. Am J Hum Genet 88: 76-82.
62. Zaitlen N, Kraft P. (2012) Heritability in the genome-wide association era. Hum Genet 131: 1655-1664.
63. Dempster ER, Lerner IM. (1950) Heritability of threshold characters. Genetics 35: 212-236.
64. Risch NJ. (2000) Searching for genetic determinants in the new millennium. Nature 405: 847-856.
65. Gazave E, Chang D, Clark AG, Keinan A. (2013) Population growth inflates the per-individual number of deleterious mutations and reduces their mean effect. Genetics 195: 969-978.
66. Haldane JBS. (1937) The effect of variation on fitness. Am Nat 71: 337-349.
67. Muller HJ. (1950) Our load of mutations. Am J Hum Genet 2: 111-176.

68. Wray NR, Purcell SM, Visscher PM. (2011) Synthetic associations created by rare variants do not explain most GWAS results. PLoS Biol 9: e1000579.
69. Visscher PM, Brown MA, McCarthy MI, Yang J. (2012) Five years of GWAS discovery. Am J Hum Genet 90: 7-24.
70. Pritchard JK. (2001) Are rare variants responsible for susceptibility to complex diseases? Am J Hum Genet 69: 124-137.
71. Thornton KR, Foran AJ, Long AD. (2013) Properties and modeling of GWAS when complex disease risk is due to non-complementing, deleterious mutations in genes of large effect. PLoS Genet 9: e1003258.
72. International Schizophrenia Consortium, Purcell SM, Wray NR, Stone JL, Visscher PM, et al. (2009) Common polygenic variation contributes to risk of schizophrenia and bipolar disorder. Nature 460: 748-752.
73. Lango Allen H, Estrada K, Lettre G, Berndt SI, Weedon MN, et al. (2010) Hundreds of variants clustered in genomic loci and biological pathways affect human height. Nature 467: 832-838.
74. Williams SM, Haines JL. (2011) Correcting away the hidden heritability. Ann Hum Genet 75: 348-350.
75. Roberts DF, Billewicz WZ, McGregor IA. (1978) Heritability of stature in a West African population. Ann Hum Genet 42: 15-24.
76. Wright A, Charlesworth B, Rudan I, Carothers A, Campbell H. (2003) A polygenic basis for late-onset disease. Trends Genet 19: 97-106.
77. Sivakumaran S, Agakov F, Theodoratou E, Prendergast JG, Zgaga L, et al. (2011) Abundant pleiotropy in human complex diseases and traits. Am J Hum Genet 89: 607-618.
78. Maher MC, Uricchio LH, Torgerson DG, Hernandez RD. (2012) Population genetics of rare variants and complex diseases. Hum Hered 74: 118-128.
79. Pavard S, Metcalf CJ. (2007) Negative selection on *BRCA1* susceptibility alleles sheds light on the population genetics of late-onset diseases and aging theory. PLoS One 2: e1206.
80. Lyman RF, Lawrence F, Nuzhdin SV, Mackay TF. (1996) Effects of single *P*-element insertions on bristle number and viability in *Drosophila melanogaster*. Genetics 143: 277-292.
81. Park JH, Gail MH, Weinberg CR, Carroll RJ, Chung CC, et al. (2011) Distribution of allele frequencies and effect sizes and their interrelationships for common genetic susceptibility variants. Proc Natl Acad Sci U S A 108: 18026-18031.
82. Torgerson DG, Boyko AR, Hernandez RD, Indap A, Hu X, et al. (2009) Evolutionary processes acting on candidate cis-regulatory regions in humans inferred from patterns of polymorphism and divergence. PLoS Genet 5: e1000592.
83. Agarwala V, Flannick J, Sunyaev S, GoT2D Consortium, Altshuler D. (2013) Evaluating empirical bounds on complex disease genetic architecture. Nat Genet 45: 1418-1427.
84. Zuk O, Schaffner SF, Samocha K, Do R, Hechter E, et al. (2014) Searching for missing heritability: Designing rare variant association studies. Proc Natl Acad Sci U S A 111: E455-64.
85. Do R, Kathiresan S, Abecasis GR. (2012) Exome sequencing and complex disease: Practical aspects of rare variant association studies. Hum Mol Genet 21: R1-9.
86. Kinnamon DD, Hershberger RE, Martin ER. (2012) Reconsidering association testing methods using single-variant test statistics as alternatives to pooling tests for sequence data with rare variants. PLoS One 7: e30238.
87. Casto AM, Feldman MW. (2011) Genome-wide association study SNPs in the human genome diversity project populations: Does selection affect unlinked SNPs with shared trait associations? PLoS Genet 7: e1001266.
88. Chen R, Corona E, Sikora M, Dudley JT, Morgan AA, et al. (2012) Type 2 diabetes risk alleles demonstrate extreme directional differentiation among human populations, compared to other diseases. PLoS Genet 8: e1002621.

89. Bustamante CD, Burchard EG, De la Vega FM. (2011) Genomics for the world. Nature 475: 163-165.
90. Carlson CS, Matise TC, North KE, Haiman CA, Fesinmeyer MD, et al. (2013) Generalization and dilution of association results from european GWAS in populations of non-European ancestry: The PAGE study. PLoS Biol 11: e1001661.
91. Shriner D, Adeyemo A, Gerry NP, Herbert A, Chen G, et al. (2009) Transferability and fine-mapping of genome-wide associated loci for adult height across human populations. PLoS One 4: e8398.
92. Waters KM, Stram DO, Hassanein MT, Le Marchand L, Wilkens LR, et al. (2010) Consistent association of type 2 diabetes risk variants found in Europeans in diverse racial and ethnic groups. PLoS Genet 6: 10.1371/journal.pgen.1001078.
93. Marigorta UM, Navarro A. (2013) High trans-ethnic replicability of GWAS results implies common causal variants. PLoS Genet 9: e1003566.

**FIGURES:**

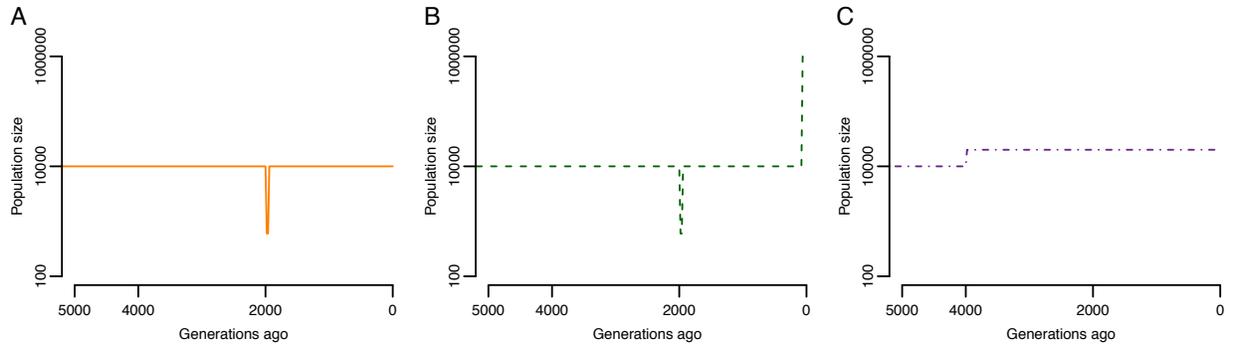

**Figure 1: Models of population size changes over time.** (**A**) A model of European population history with a severe bottleneck starting 2000 generations ago (BN). (**B**) A similar model of European population history shown in (**A**), except that here the population instantaneously expanded 100-fold 80 generations ago (BN+growth). (**C**) A model with a 2-fold ancient expansion (Old growth). This is a possible model for African population history.

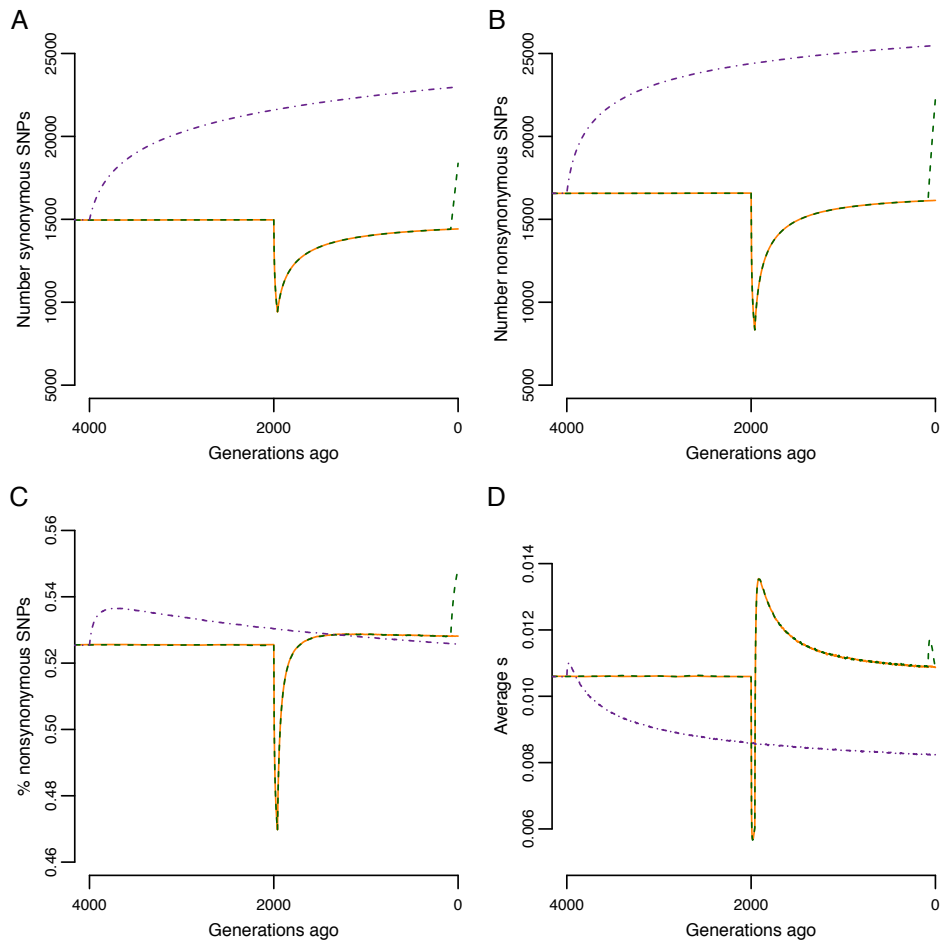

**Figure 2: Changes in genetic variation over time as a function of population size.** Solid orange lines denote the bottlenecked population that did not recently expand (BN). Dashed green lines denote a population that expanded 80 generations ago (BN+growth). Note that the lines from the two populations overlap except in the last 80 generations. Dashed purple lines denote the population that underwent an ancient expansion (Old growth). (**A**) Number of synonymous SNPs segregating in the sample. (**B**) Number of nonsynonymous SNPs segregating in the sample. (**C**) Proportion of SNPs segregating in the sample that are nonsynonymous. (**D**) Average fitness effects of nonsynonymous SNPs segregating in the sample. Samples of 1000 chromosomes were taken at different time points throughout the simulation. Results are averaged over 1000 simulation replicates.

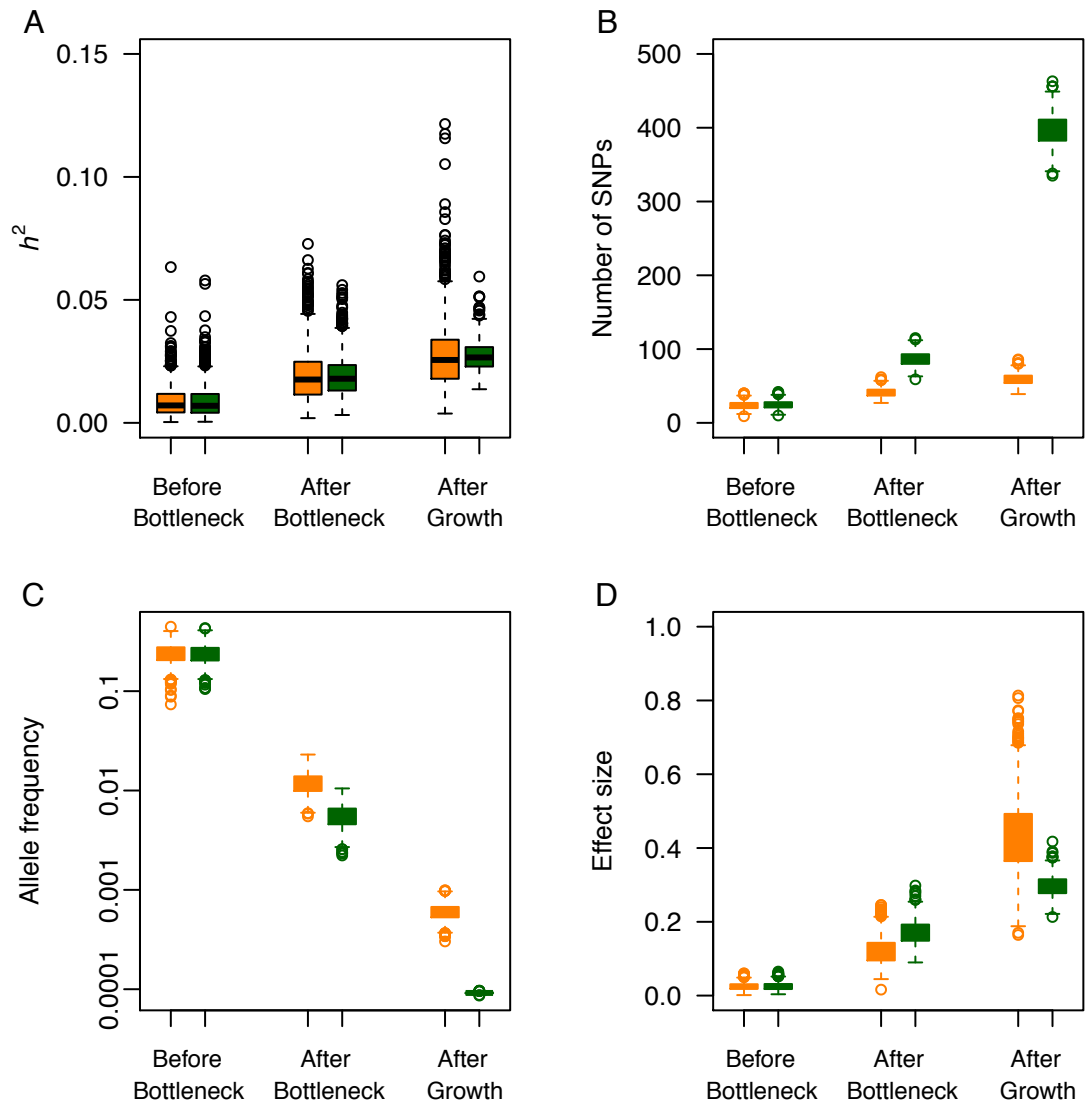

**Figure 3: Effect of recent population growth on the heritability attributable to mutations of different ages when τ = 0.5.** Orange boxes denote the bottlenecked population that did not recently expand (BN). Green boxes denote a population that expanded 80 generations ago (BN+growth). "Before bottleneck" refers to mutations that arose more than 1960 generations ago (before or during the bottleneck). "After bottleneck" refers to mutations that arose after the population recovered from the bottleneck, but earlier than 80 generations ago. "After growth" refers to mutations that arose within the last 80 generations (after the population expanded). **(A)**

Heritability attributed to mutations of different ages. Note that recent population growth does not affect the median heritability attributable to mutations of different ages. **(B)** Number of SNPs segregating in the present-day that arose during the different time intervals. **(C)** Mean allele frequency of SNPs that are segregating in the present-day that arose during the different time intervals. **(D)** Mean effect size of SNPs that are segregating in the present-day that arose during the different time intervals. Here $h_C^2 = 0.05$ and $M = 70$ kb.

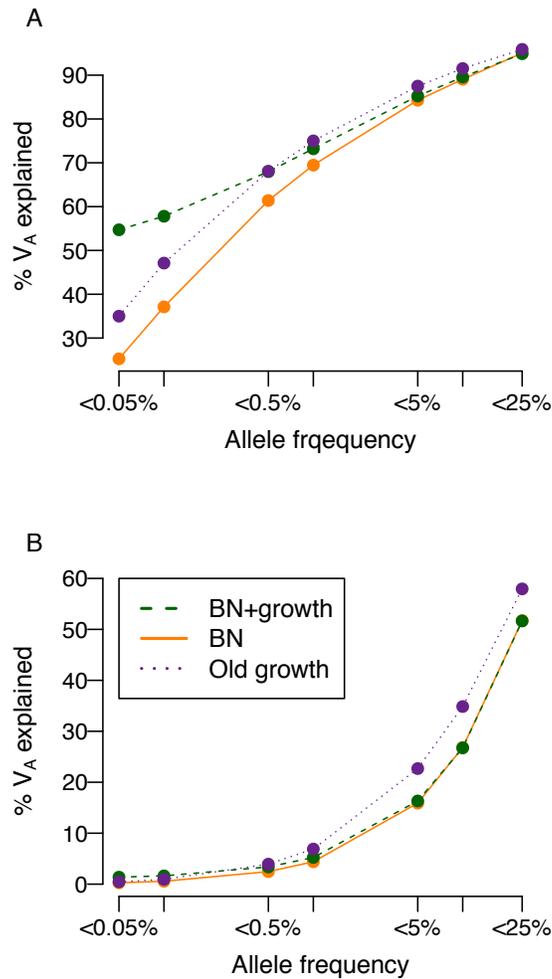

**Figure 4: Cumulative distribution of the amount of the additive genetic variance of a trait ($V_A$; *y*-axis) explained by SNPs segregating below a given frequency in the population (*x*-axis).** (**A**) A SNP's effect on the trait is correlated with its effect on fitness ($\tau = 0.5$). Note that the population that experienced recent growth (green; BN+growth) has a higher proportion of $V_A$ accounted for by low-frequency SNPs (<0.1% frequency) than the populations that did not recently expand (orange and purple; BN and Old growth). (**B**) A SNP's effect on the trait is independent of its effect on fitness ($\tau = 0$). Note that less of $V_A$ is accounted for by low-frequency variants than when the trait is correlated with fitness (**A**). Here $h_C^2 = 0.05$ and $M = 70$ kb.

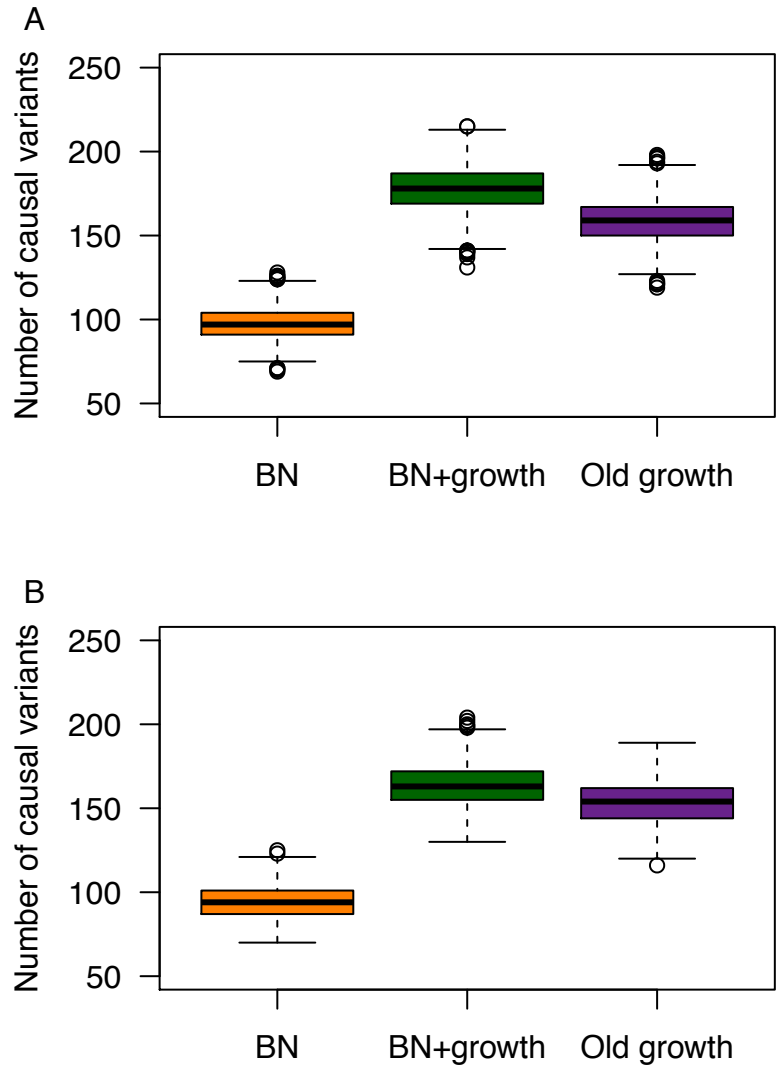

**Figure 5: The number of causal mutations in a sample of 1000 cases from each simulated population.** (**A**) A SNP's effect on the trait is correlated with its effect on fitness ($\tau = 0.5$). Note that the population that experienced recent growth (green; BN+growth) has a higher number of causal mutations than the populations that did not recently expand (orange; BN). (**B**) A SNP's effect on the trait is independent of its effect on fitness ($\tau = 0$). Here $h_C^2 = 0.05$ and $M = 70$ kb.

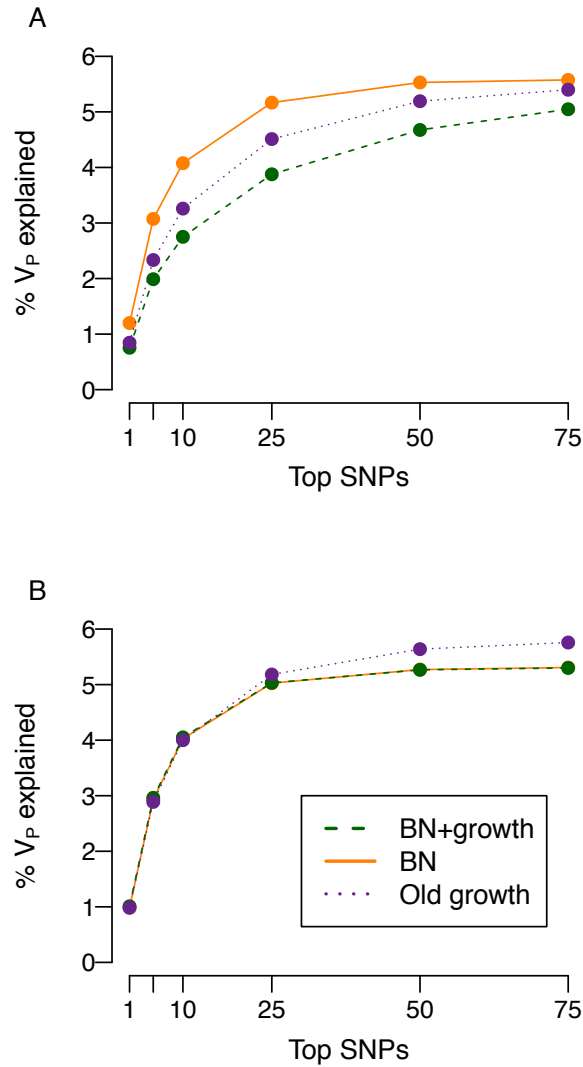

**Figure 6: Cumulative distribution of the amount of the phenotypic variance of a trait ($V_P$; y-axis) explained by the SNPs that explain the most variance (*x*-axis).** (**A**) A SNP's effect on the trait is correlated with its effect on fitness ($\tau = 0.5$). Note that the population that experienced recent growth (green; BN+growth) has a lower proportion of $V_P$ accounted for by the top SNPs that account for most of the variance than the populations that did not recently expand (orange and purple; BN and Old growth). (**B**) A SNP's effect on the trait is independent of its effect on fitness ($\tau = 0$). Here $h_C^2 = 0.05$ and $M = 70$ kb.

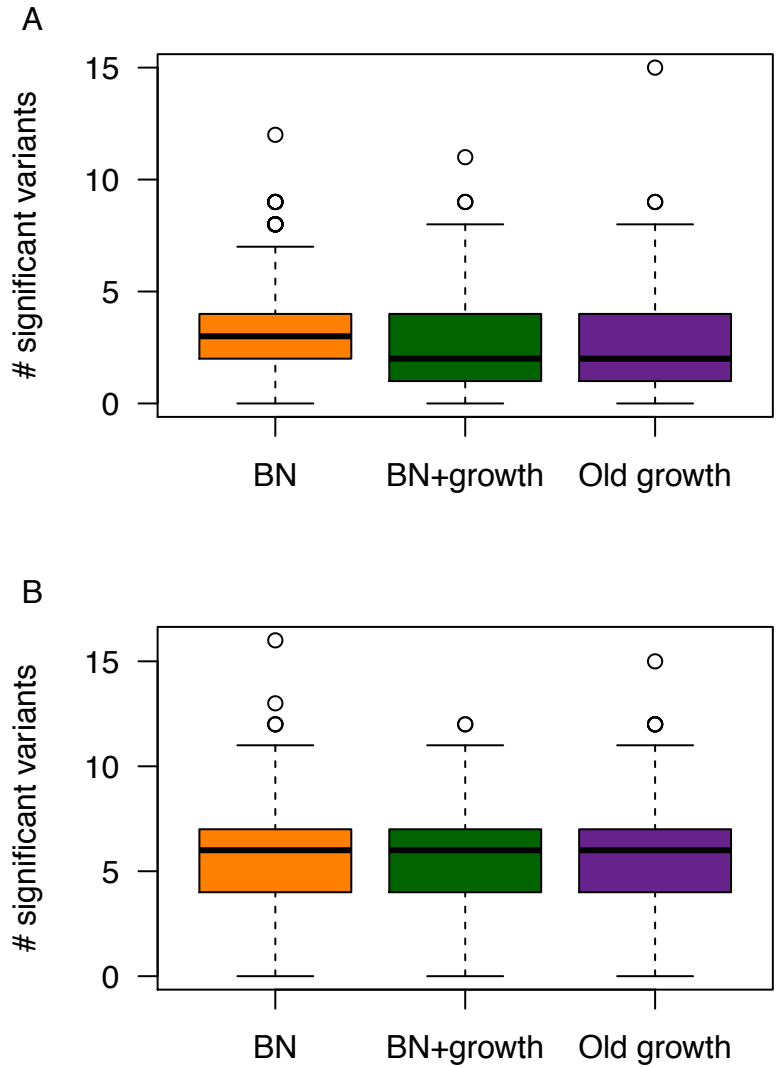

**Figure 7: The number of causal SNPs with a significant *P*-value (<1 x 10⁻⁵) in the single-marker association test for different models of population history.** (**A**) A SNP's effect on the trait is correlated with its effect on fitness (τ = 0.5). (**B**) A SNP's effect on the trait is independent of its effect on fitness (τ = 0). Here $h_C^2 = 0.3$ and *M* = 70 kb.

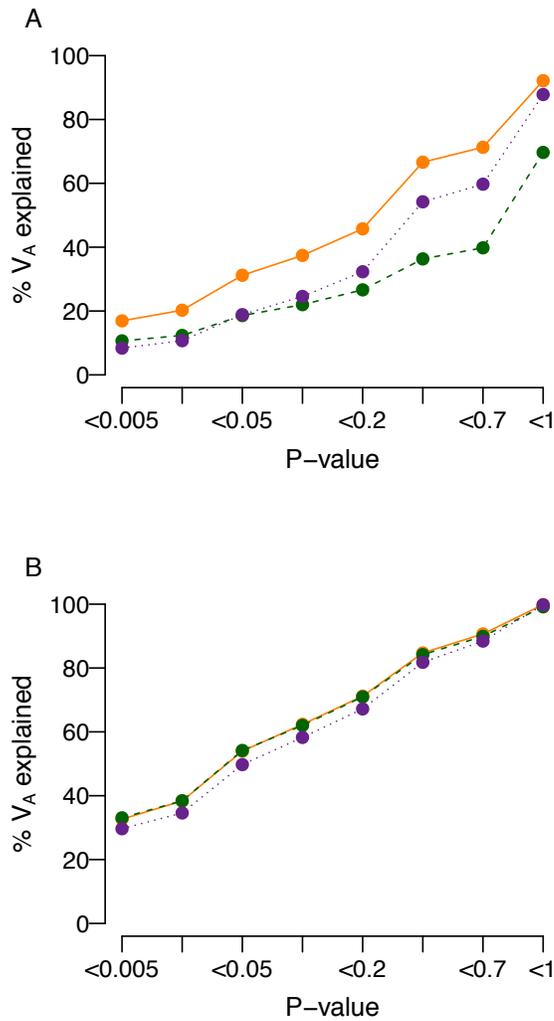

**Figure 8: Cumulative distribution of the amount of the additive genetic variance of a trait ($V_A$; y-axis) explained by SNPs with a given single-marker association test *P*-value (x-axis).** (**A**) A SNP's effect on the trait is correlated with its effect on fitness ($\tau = 0.5$). Note that the population that experienced recent growth (green line; BN+growth) has a lower proportion of $V_A$ accounted for by SNPs at any *P*-value threshold than the populations that did not recently expand (orange and purple lines; BN and Old growth). (**B**) A SNP's effect on the trait is independent of its effect on fitness ($\tau = 0$). Here note that the SNPs with low *P*-values (<0.05) account for most of $V_A$ regardless of the demographic history of the population. Here $h_C^2 = 0.05$ and $M = 70$ kb.

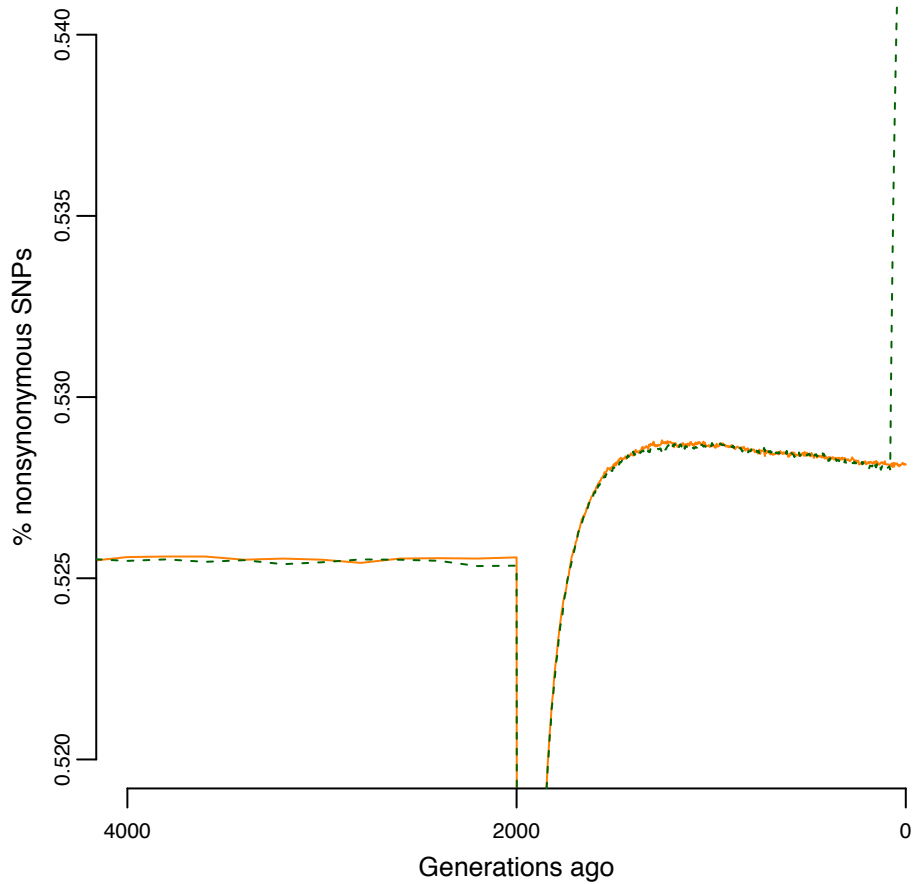

**Figure S1: Proportion of nonsynonymous SNPs over time.** Note that the proportion of nonsynonymous SNPs after the bottleneck (near 0 generations) is higher than that in the ancestral population (at time 4000 generations ago) for both the model with recent population growth (dashed green line) and the model without recent population growth (solid orange line).

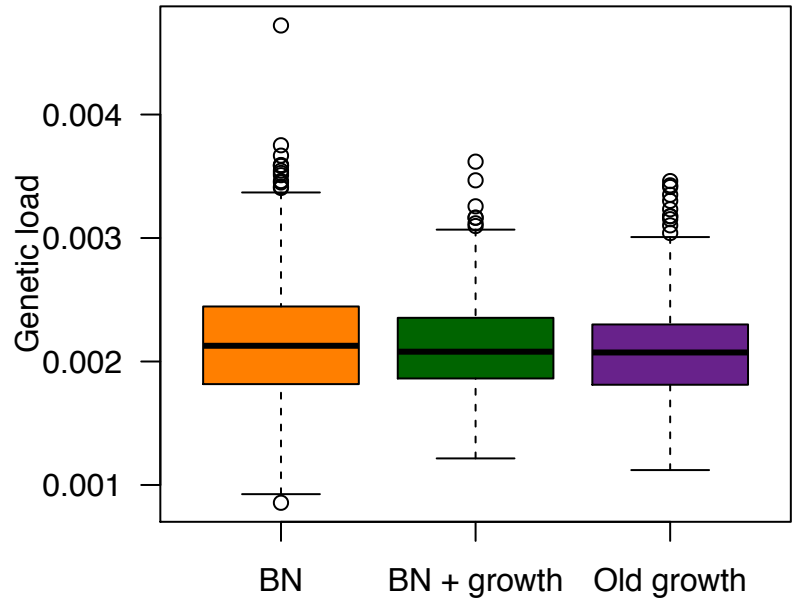

**Figure S2: Population history has little effect on the genetic load.** Genetic load was calculated for all SNPs segregating in a sample of 6,000 individuals taken from each demographic history.

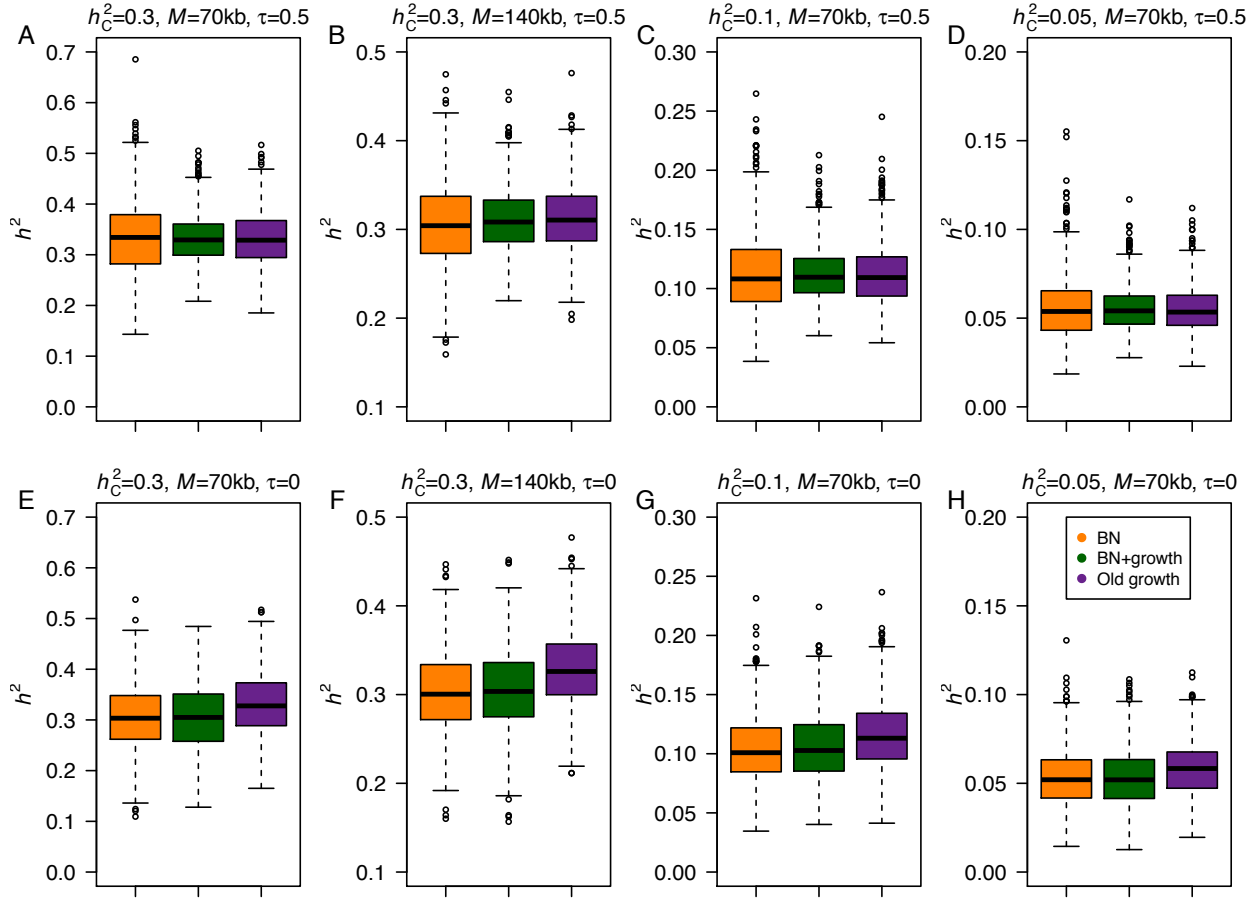

**Figure S3: Population history has little effect on the narrow-sense heritability ($h^2$) of a trait.** (**A-D**) A SNP's effect on the trait is correlated with its effect on fitness ($\tau = 0.5$). (**E-H**) A SNP's effect on the trait is independent of its effect on fitness ($\tau = 0$). (**A, E**) $h_C^2 = 0.3$ and $M = 70$ kb. (**B, F**) $h_C^2 = 0.3$ and $M = 140$ kb. (**C, G**) $h_C^2 = 0.1$ and $M = 70$ kb. (**D, G**) $h_C^2 = 0.05$ and $M = 70$ kb. Narrow sense heritability was computed for each demographic model as $h^2 = \dfrac{V_A}{V_A + V_E}$. Here $V_E = 1 - h_C^2$ for all scenarios. $V_A = \sum\limits_{i \text{ SNPs}} 2 p_i (1 - p_i) \alpha_i^2$, $p_i$ is the frequency of the $i^{\text{th}}$ SNP, and $\alpha_i$ is the $i^{\text{th}}$ SNP's effect on the trait.

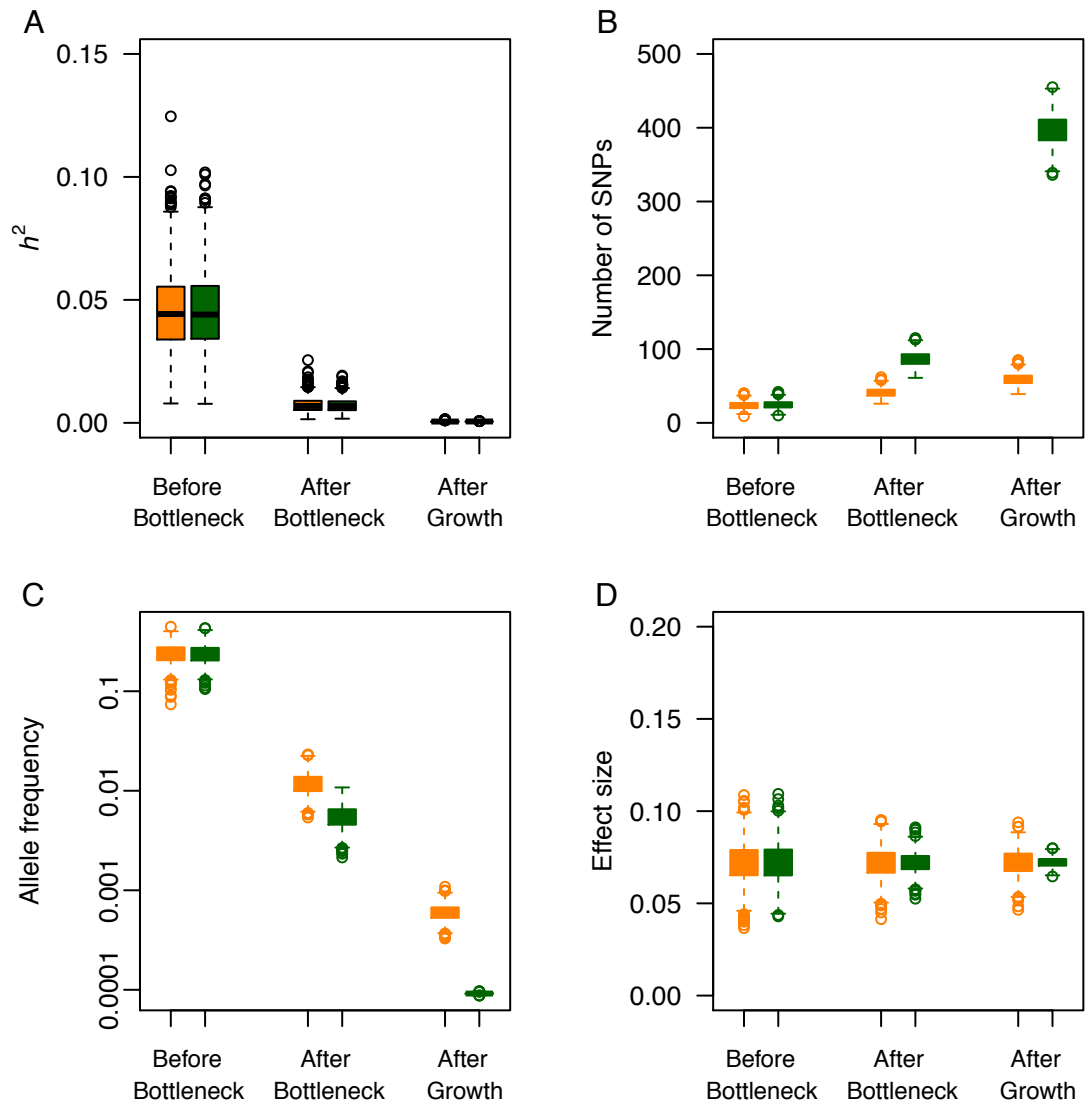

**Figure S4: Effect of recent population growth on the heritability attributable to mutations of different ages when τ = 0.** Orange boxes denote the bottlenecked population that did not recently expand (BN). Green boxes denote a population that expanded 80 generations ago (BN+growth). "Before bottleneck" refers to mutations that arose more than 1960 generations ago (before or during the bottleneck). "After bottleneck" refers to mutations that arose after the population recovered from the bottleneck, but earlier than 80 generations ago. "After growth" refers to mutations that arose within the last 80 generations (after the population expanded). **(A)**

Heritability attributed to mutations of different ages. Note that recent population growth does not affect the median heritability attributable to mutations of different ages. **(B)** Number of SNPs segregating in the present-day that arose during the different time intervals. **(C)** Mean allele frequency of SNPs that are segregating in the present-day that arose during the different time intervals. **(D)** Mean effect size of SNPs that are segregating in the present-day that arose during the different time intervals. Here $h_C^2 = 0.05$ and $M = 70$ kb.

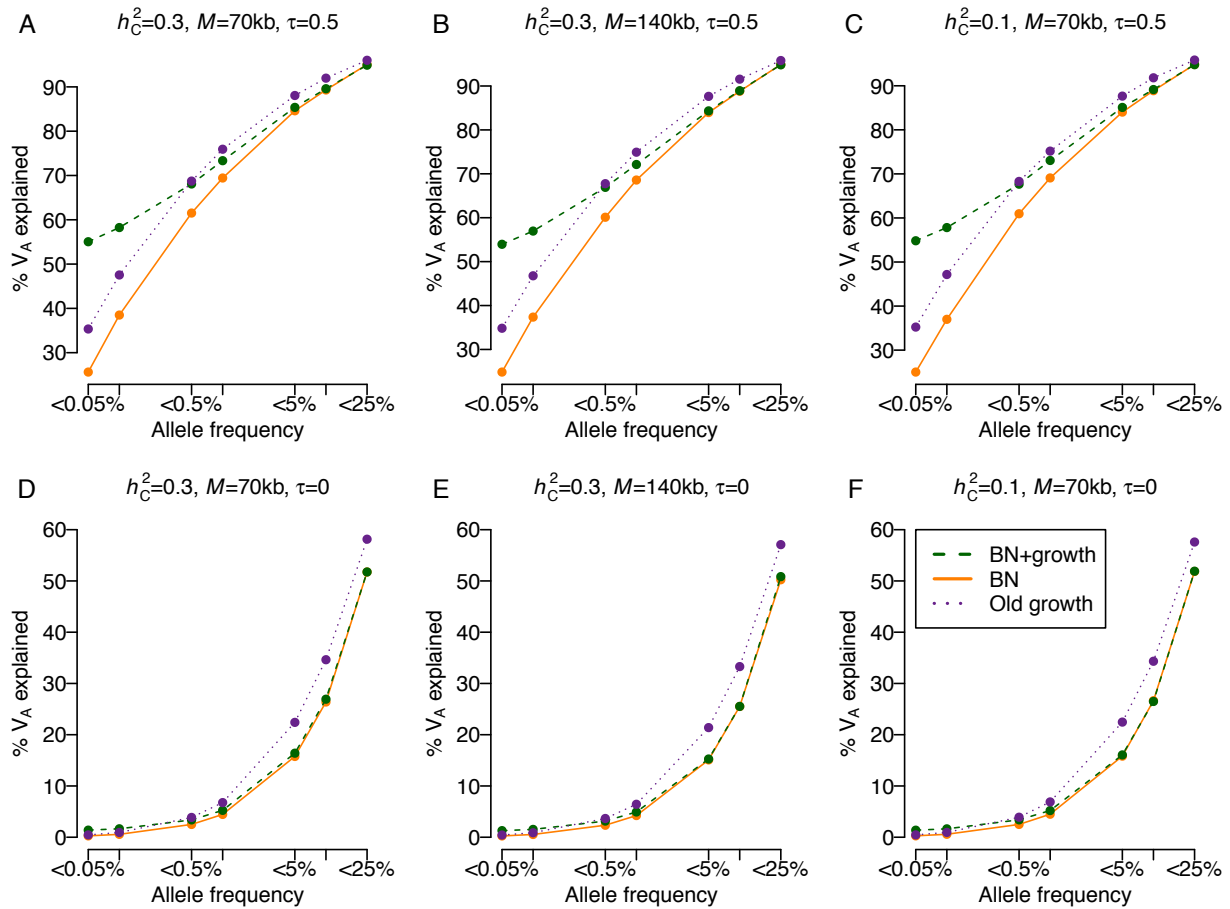

**Figure S5: Cumulative distribution of the amount of the additive genetic variance of a trait ($V_A$; *y*-axis) explained by SNPs segregating below a given frequency in the population (*x*-axis) for additional models of the trait.** (**A-C**) A SNP's effect on the trait is correlated with its effect on fitness ($\tau = 0.5$). Note that the population that experienced recent growth (green; BN+growth) has a higher proportion of $V_A$ accounted for by low-frequency SNPs (<0.1% frequency) than the populations that did not recently expand (orange and purple; BN and Old growth). (**D-F**) A SNP's effect on the trait is independent of its effect on fitness ($\tau = 0$). Note that less of $V_A$ is accounted for by low-frequency variants than when the trait is correlated with fitness (**A**).

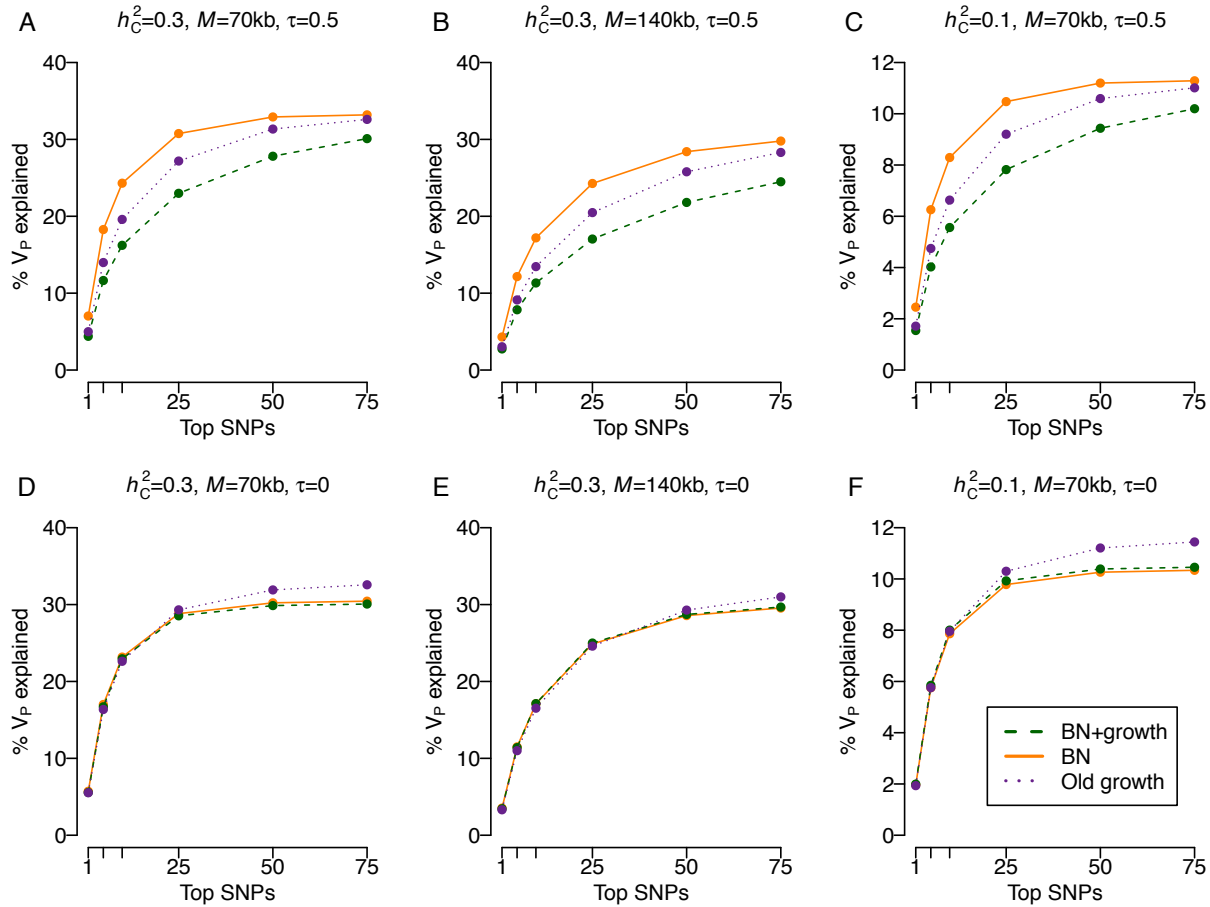

**Figure S6: Cumulative distribution of the amount of the phenotypic variance of a trait ($V_P$; y-axis) explained by the SNPs that explain the most variance (*x*-axis) for additional models of the trait.** (**A-C**) A SNP's effect on the trait is correlated with its effect on fitness ($\tau = 0.5$). Note that the population that experienced recent growth (green; BN+growth) has a lower proportion of $V_P$ accounted for by the top SNPs that account for most of the variance than the populations that did not recently expand (orange and purple; BN and Old growth). (**D-F**) A SNP's effect on the trait is independent of its effect on fitness ($\tau = 0$).

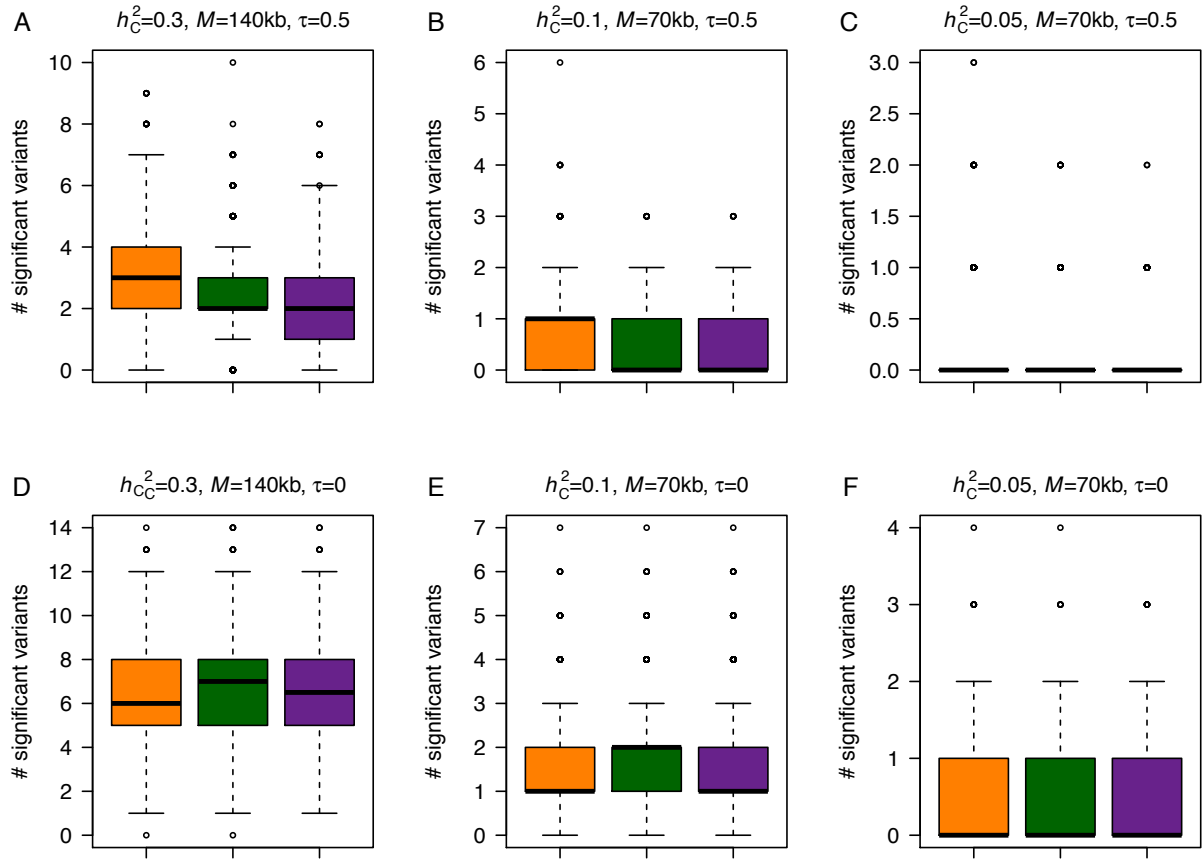

**Figure S7: The number of causal SNPs with a significant *P*-value in the single-marker association test for additional models of the trait.** Orange denotes the bottleneck demographic model (BN). Green denotes the bottleneck and recent growth model (BN+growth). Purple denotes the ancient growth model (Old growth). (**A-C**) A SNP's effect on the trait is correlated with its effect on fitness ($\tau = 0.5$). (**D-F**) A SNP's effect on the trait is independent of its effect on fitness ($\tau = 0$). (**A, D**) $h_C^2 = 0.3$ and $M = 140$ kb. (**B, E**) $h_C^2 = 0.1$ and $M = 70$ kb. (**C, F**) $h_C^2 = 0.05$ and $M = 70$ kb.

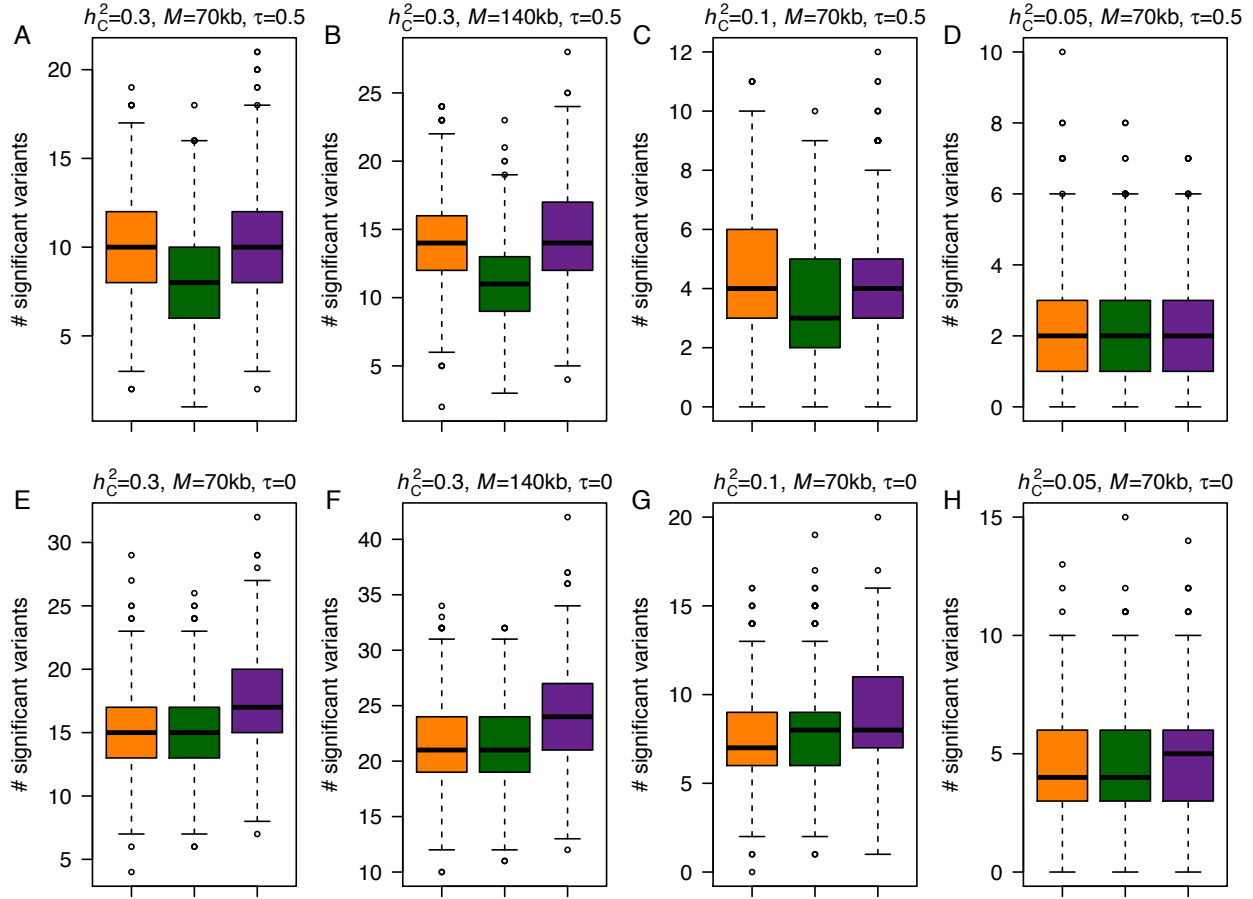

**Figure S8: The number of causal SNPs with a *P*-value <1 x 10$^{-2}$ in the single-marker association test for different models of population history and the trait.** Orange denotes the bottleneck demographic model (BN). Green denotes the bottleneck and recent growth model (BN+growth). Purple denotes the ancient growth model (Old growth). (**A-D**) A SNP's effect on the trait is correlated with its effect on fitness ($\tau = 0.5$). (**E-H**) A SNP's effect on the trait is independent of its effect on fitness ($\tau = 0$). (**A, E**) $h_C^2 = 0.3$ and $M = 70$ kb. (**B, F**) $h_C^2 = 0.3$ and $M = 140$ kb. (**C, G**) $h_C^2 = 0.1$ and $M = 70$ kb. (**D, G**) $h_C^2 = 0.05$ and $M = 70$ kb.

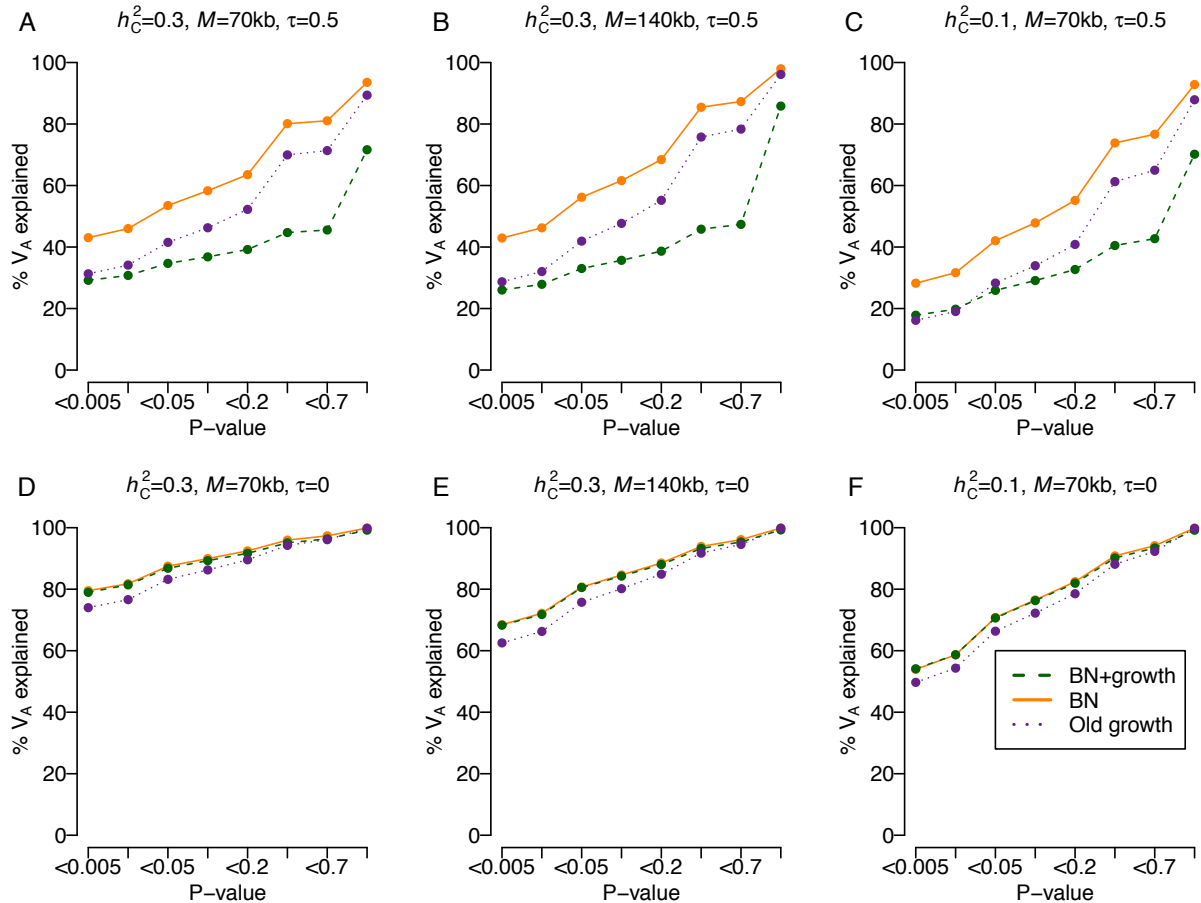

**Figure S9: Cumulative distribution of the amount of the additive genetic variance of a trait ($V_A$; *y*-axis) explained by SNPs with a given single-marker association test *P*-value (*x*-axis) for additional models of the trait.** (**A-C**) A SNP's effect on the trait is correlated with its effect on fitness ($\tau = 0.5$). Note that the population that experienced recent growth (green line; BN+growth) has a lower proportion of $V_A$ accounted for by SNPs at any *P*-value threshold than the populations that did not recently expand (orange and purple lines; BN and Old growth). (**D-F**) A SNP's effect on the trait is independent of its effect on fitness ($\tau = 0$). Note that the SNPs with low *P*-values (<0.05) account for most of $V_A$ regardless of the demographic history of the population. (**A, D**) $h_C^2 = 0.3$ and $M = 70$ kb. (**B, E**) $h_C^2 = 0.3$ and $M = 140$ kb. (**C, F**) $h_C^2 = 0.1$ and $M = 70$ kb.

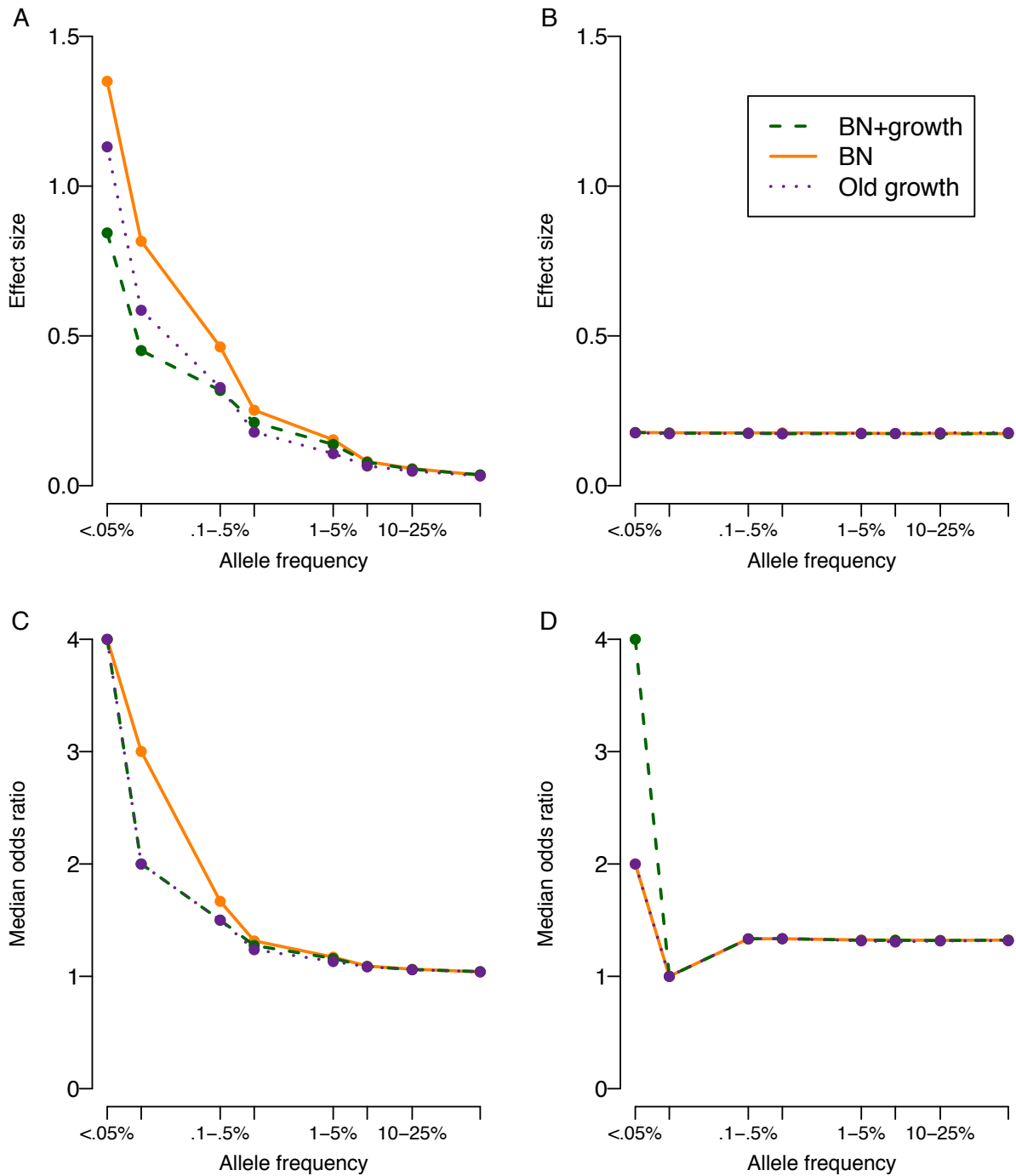

**Figure S10: The relationship between a mutation's effect on the trait and its allele frequency.** Statistics were calculated for each simulation replicate and then averaged over the 1000 simulation replicates. **(A, C)** A SNP's effect on the trait is correlated with its effect on fitness (τ = 0.5). **(B, D)** A SNP's effect on the trait is independent of its effect on fitness (τ = 0). **(A-B)** Average effect size (on the liability scale) for SNPs having the allele frequency (in the population) specified on the *x*-axis. **(C-D)** Median odds ratios (ORs) computed from a sample of 1000 cases and controls across all SNPs in a simulation replicate having the allele frequency (in the population) specified on the *x*-axis. Note, median ORs equal to infinity (due to many case-only variants) were set to 4 for plotting purposes Here $h_C^2 = 0.3$ and *M* = 70 kb.

Table S1: Values of the constant $C$ used to generate the desired heritability for different values of $h_C^2$ and $M$.

| $\tau$ | $h_C^2 = 0.3$; $M = 70$ kb | $h_C^2 = 0.3$; $M = 140$ kb | $h_C^2 = 0.1$; $M = 70$ kb | $h_C^2 = 0.05$; $M = 70$ kb |
|---|---|---|---|---|
| 0 | 0.03058 | 0.014891 | 0.0104 | 0.005219 |
| 0.5 | 60 | 26.69925 | 19.24 | 9.435 |

$\tau$ denotes the relationship between a mutation's effect on fitness and the trait. $h_C^2$ refers to the heritability that the simulation was calibrated to in a constant size population. Note, the same value of $C$ was used under all models of population history (see Methods).

Table S2: Average percentage of the phenotype variance ($V_P$) explained by the top 50 SNPs that explain the most variance under different models of population history, $h_C^2$ and $M$.

| $\tau$ | Population | $h_C^2 = 0.3$; $M$ = 70 kb | $h_C^2 = 0.3$; $M$ = 140 kb | $h_C^2 = 0.1$; $M$ = 70 kb | $h_C^2 = 0.05$; $M$ = 70 kb |
|---|---|---|---|---|---|
| 0 | BN+growth | 29.9 | 28.7 | 10.3 | 5.3 |
|   | BN | 30.2 | 28.6 | 10.3 | 5.3 |
|   | Old growth | 31.9 | 29.3 | 11.2 | 5.6 |
| 0.5 | BN+growth | 27.8 | 21.8 | 9.4 | 4.7 |
|   | BN | 32.9 | 28.4 | 11.2 | 5.5 |
|   | Old growth | 31.3 | 25.6 | 10.6 | 5.2 |

$\tau$ denotes the relationship between a mutation's effect on fitness and the trait. $h_C^2$ refers to the heritability that the simulation was calibrated to in a constant size population. For this reason, some of the observed averages of $V_P$ may be greater than $h_C^2$ listed in the column heading. $M$ refers to the mutational target size (see Methods). Note that certain models predict that the top 50 SNPs, in aggregate, explain <10% of the $V_P$.

Table S3: Average number of GWAS hits expected in samples of 1,000 cases and 1,000 controls under different models of population history, $h_C^2$ and $M$.

| τ | Population | $h_C^2 = 0.3$; $M = 70$ kb | $h_C^2 = 0.3$; $M = 140$ kb | $h_C^2 = 0.1$; $M = 70$ kb | $h_C^2 = 0.05$; $M = 70$ kb |
|---|---|---|---|---|---|
| 0 | BN+growth | 3.9 | 3.3 | 0.6 | 0.1 |
|   | BN | 4.0 | 3.2 | 0.6 | 0.1 |
|   | Old growth | 3.9 | 3.1 | 0.6 | 0.1 |
| 0.5 | BN+growth | 1.4 | 1.2 | 0.2 | 0.0 |
|   | BN | 1.7 | 1.7 | 0.2 | 0.0 |
|   | Old growth | 1.1 | 0.88 | 0.1 | 0.0 |

τ denotes the relationship between a mutation's effect on fitness and the trait. $h_C^2$ refers to the heritability that the simulation was calibrated to in a constant size population. A significance threshold of $5 \times 10^{-8}$ was used.

**Supplementary Text S1**

**Additional results on recent growth and deleterious variation**

*Looking into the future: re-attaining equilibrium*

Very recent population growth leads to an increase in the proportion of nonsynonymous SNPs in the population compared to a population that has not recently expanded. Thus, the recent population growth has pushed the population out of equilibrium. But, as natural selection eliminates the new deleterious mutations from the population, the proportion of nonsynonymous SNPs in the population and the average deleterious effect of a SNP will decrease. Eventually, assuming no further demographic changes, the proportion of nonsynonymous SNPs in the population will attain the new equilibrium value for the larger population size (**Fig. S1.1**).

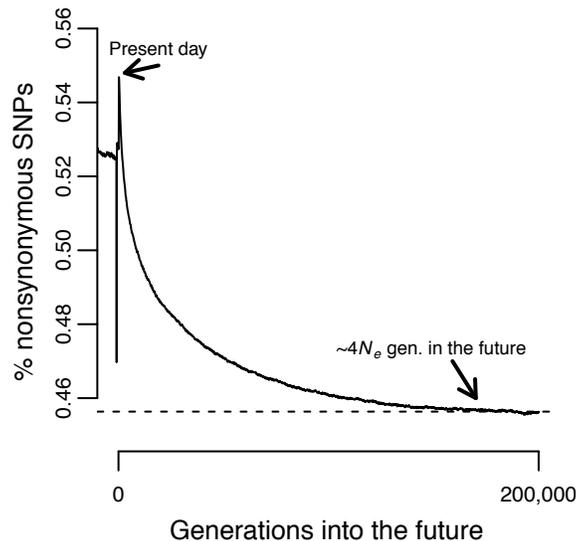

**Fig. S1.1: Figure S5: Re-attaining the equilibrium value for the proportion of nonsynonymous SNPs after a 5-fold population expansion**. The population instantaneously expanded 5-fold at time 0. Solid curve shows how the proportion of nonsynonymous SNPs is expected to change over time. The dashed line denotes the proportion of nonsynonymous SNPs at equilibrium for the larger population size. After approximately $4N_e$ (where $N_e$ is the size after the expansion) generations, the proportion of nonsynonymous SNPs in the expanded population reaches the equilibrium proportion for the larger population size.

How long will it take for human populations to reach this new equilibrium? By running the simulations with 5-fold growth for many generations into the future, it will take roughly 200,000 or $4N_e$ generations for the proportion of nonsynonymous SNPs segregating in the population to decrease to the new equilibrium value for the larger population size (**Fig. S1.1**). This agrees well with the classic result that, conditional on fixation, a new neutral mutation takes roughly $4N_e$ generations to become fixed in the population [1]. Thus, extrapolating to 100-fold growth, it would take roughly 4 million generations to reach the equilibrium proportion of nonsynonymous SNPs in the population.

*Effect of different growth rates*

I also examine the effect that different magnitudes of population growth have on the proportion of nonsynonymous SNPs in the population (**Fig. S1.2A**) and the average fitness effects of segregating deleterious variants (**Fig. S1.2B**). As expected, stronger recent population growth results in a higher proportion of nonsynonymous SNPs in the population (**Fig. S1.2A**). Interestingly, the difference between 10-fold growth and 100-fold growth appears rather slight compared to the difference between 0 and 5-fold growth. This result suggests that even a small amount of recent growth may be sufficient to affect patterns of weakly deleterious mutations. The picture for the average fitness effects of segregating deleterious mutations is more complex (**Fig. S1.2B**). Here, as the magnitude of growth increases, the average segregating SNP becomes less deleterious. This effect can be explained by selection being more efficient in the larger population (as suggested in a recent paper by Gazave et al. [2]). Interestingly, the 80 generations since the expansion has been sufficient time for selection to have begun removing some of the most deleterious mutations (**Fig. S1.3).**

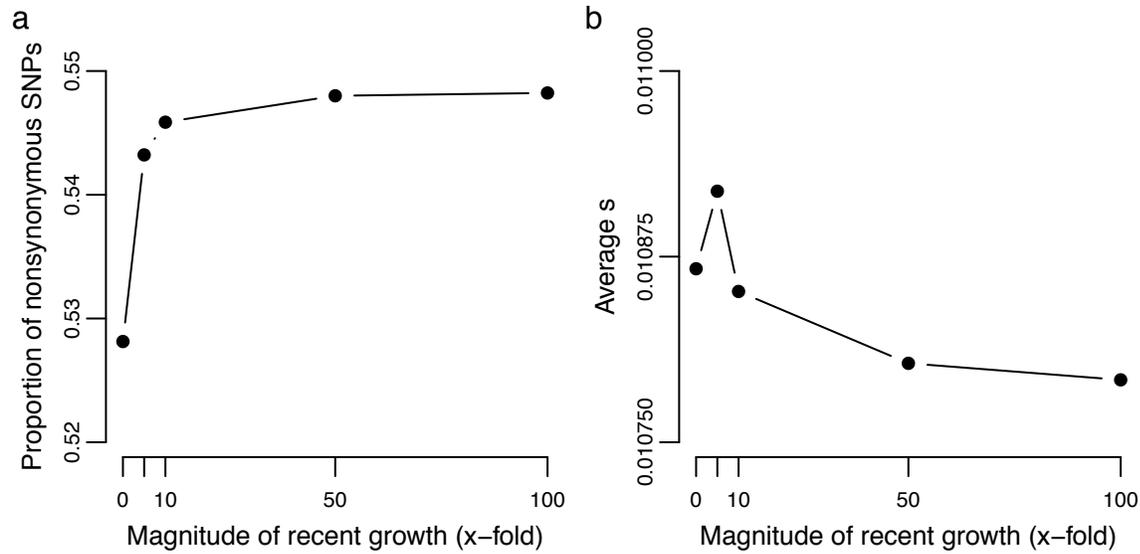

**Figure S1.2: Changes in genetic variation as a function of the magnitude of recent population growth.** In all cases, populations instantaneously expanded by the factor shown on the *x*-axis 80 generations ago. (**A**) Proportion of SNPs segregating in the sample that are nonsynonymous. (**B**) Average fitness effects of nonsynonymous SNPs segregating in the sample. Samples of 1000 chromosomes were taken at the simulation, reflecting the patterns in the present day. Results are averaged over 1000 simulation replicates.

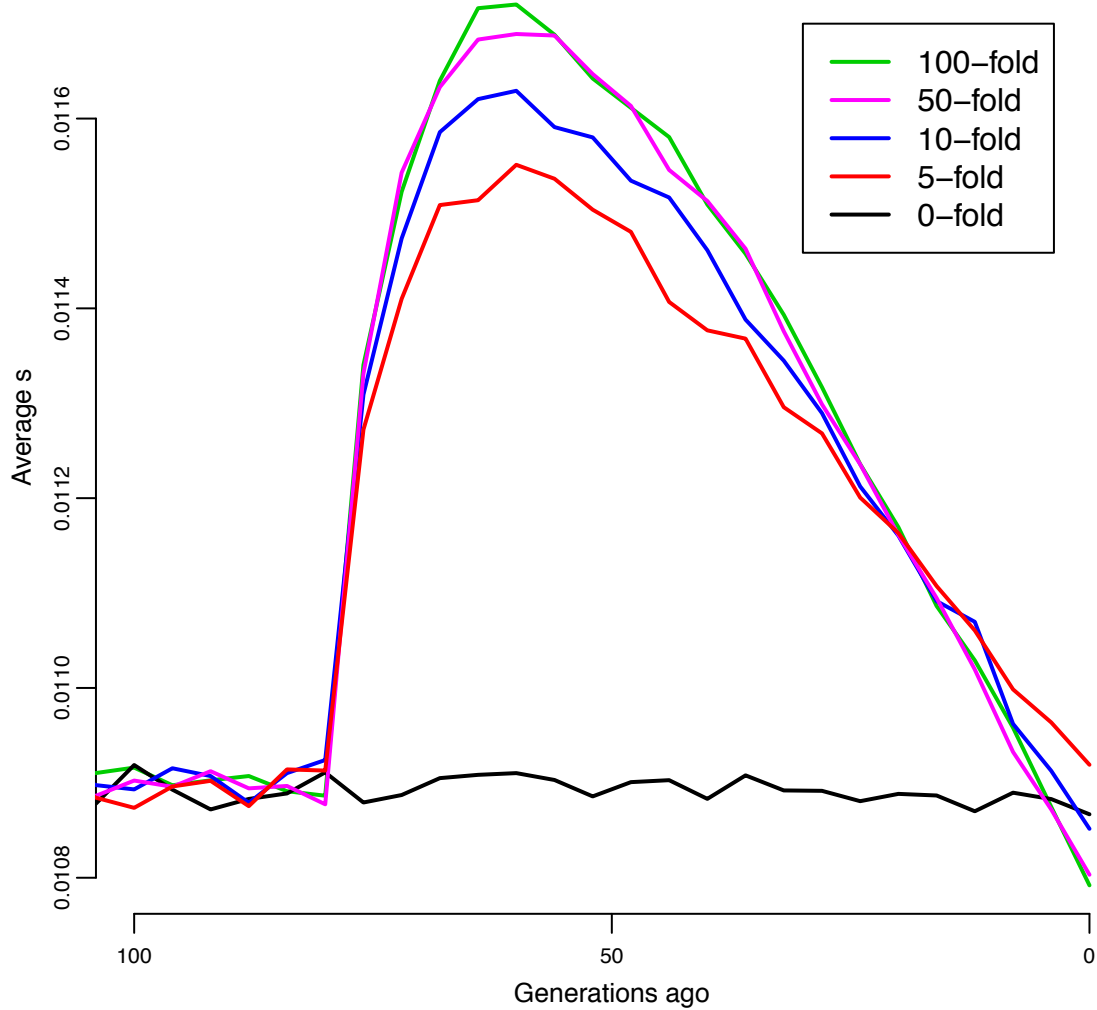

**Figure S1.3: Average fitness effect of a nonsynonymous mutation in populations that have undergone various amounts of growth in the last 80 generations.** Note that at 50 generations ago, the average SNP was most deleterious in the populations that experienced the greatest expansion (100-fold and 50-fold). This is due to the increased input of new deleterious mutations as a result of the population expansion. However, in the present day (0 generations ago), the average SNP is least deleterious in the populations that expanded the most. This is due to the increased efficacy of purifying selection in a large population. Thus, many of the most deleterious mutations are quickly eliminated from the population.

# REFERNECES

1. Kimura M, Ohta T. (1969) The average number of generations until fixation of a mutant gene in a finite population. Genetics 61: 763-771.
2. Gazave E, Chang D, Clark AG, Keinan A. (2013) Population growth inflates the per-individual number of deleterious mutations and reduces their mean effect. Genetics 195: 969-978.

**Supplementary Text S2**

**Additional results on the effect of demography on the power of association tests**

All results described here were obtained using $h_C^2 = 0.3$, $M = 70$ kb, and a critical value for Fisher's exactly test of 1 x 10$^{-5}$.

*Power as a function of $V_A$*

I examine the power to detect an association as a function of the amount of the $V_A$ that a given SNP explains (**Fig. S2.1**). When a mutation's effect on fitness is correlated with its effect on the trait ($\tau = 0.5$), power to detect an association with a SNP that explains much of the $V_A$ (>5%) is higher in the population that recently expanded (BN+growth; green lines in **Fig. S2.1A**) than a population that did not (BN; orange line in **Fig. S2.1A**). This increase in power comes from the fact that those SNPs that explain a lot of $V_A$ in the recently expanded population tend to have smaller effect sizes than those SNPs in the population that did not recently expand (**Fig. S2.1C**). Because, under this model, the effect size of a mutation is correlated with the strength of selection acting against it (**Fig. S10**), those SNPs that have smaller effect sizes are also less deleterious and can reach higher frequency in the population. This pattern is demonstrated in **Fig. S2.1E**, which shows that the SNPs that each account for about 5% of the $V_A$ in the BN+growth population have a higher average frequency than those SNPs that explain a similar proportion of $V_A$ in the BN population. This increase in allele frequency results in an increase in power for the single-marker association test. An obvious question is why do the SNPs that explain >5% of the $V_A$ in the BN+growth population have smaller effect sizes than SNPs that explain similar amounts of $V_A$ in the BN population? This counter-intuitive pattern can be explained by noting that in order for a given SNP to explain >5% of the $V_A$, it must be relatively common, or have a large effect, or both. In the BN population, there are many low-frequency mutations of strong effect which are moderately deleterious that may explain >5% of the $V_A$. In the BN+growth population, however, many of these low-frequency deleterious mutations with strong effects are eliminated from the population due to the increased efficacy of purifying selection in the large population. Those mutations that are left behind in the recently expanded population that can explain >5% of the $V_A$ will tend to be less deleterious, have smaller effect sizes (**Fig. S2.1C**), but higher allele frequencies as compared to the mutations that explain >5% of the $V_A$ in the BN population (**Fig. S2.1E**). The recent population growth also increases the number of new deleterious mutations with large effect sizes that enter the population. However, these mutations are so rare that they are unlikely to contribute >5% of the $V_A$, and so they are not relevant for the present discussion.

When a mutation's effect on fitness is not correlated with its effect on the trait ($\tau = 0$), I find that demography has little effect on the power of the association test (**Fig. S2.1B**). For all three models of population history, power to detect mutations that explain more of the $V_A$ is high. Further, the SNPs that tend to explain most of the $V_A$ tend to be at higher allele frequency (**Fig. S2.1D**) and have larger effect sizes than those mutations that explain less $V_A$ (**Fig. S2.1F**).

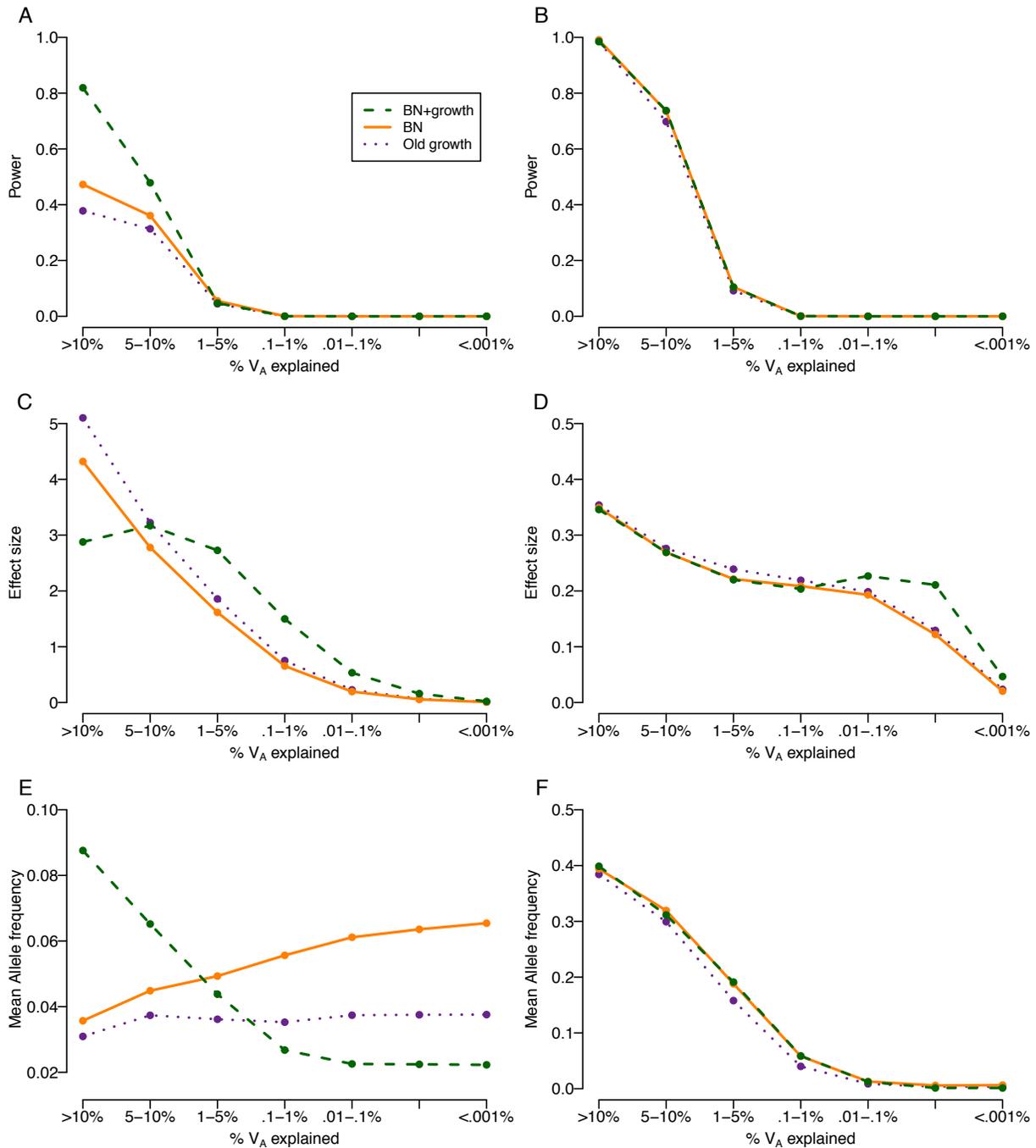

**Figure S2.1: Properties of single-marker association tests for causal variants in a study of 1000 cases and 1000 controls, binned by the additive genetic variance explained by the SNP ($V_A$; *x*-axis).** **(A, C, E)** A SNP's effect on the trait is correlated with its effect on fitness ($\tau = 0.5$). **(B, D, F)** A SNP's effect on the trait is independent of its effect on fitness ($\tau = 0$). **(A-B)** Power. **(C-D)** Average effect size of the causal variant on the liability scale. **(E-F)** Average allele frequency of the causal variants.

*Power to detect the SNPs that explain the most $V_A$*

I also investigate the power to detect the top SNPs that explain the most $V_A$ in each simulation replicate (**Fig. S2.2**). Similar trends to those discussed above apply here as well. For example, when $\tau = 0.5$, the power to detect the SNP that explains the most $V_A$ in a given simulation replicate is 46% in the BN population (orange line in **Fig. S2.2A**). Power to detect the top SNP in the BN+growth population is considerably higher—above 70%. This is due to the top SNPs being at higher allele frequency in the expanded population than in the non-expanded population (**Fig S2.2E**), increasing power. Further, the top SNPs are less likely to be private to cases in the recently expanded population than in the population that did not expand (**Fig. S2.2C**). This pattern supports the idea that the top SNPs that explain the most $V_A$ are at higher frequency in the expanded population. Finally, as noted above, when a mutation's effect on fitness is not correlated with its effect on the trait ($\tau = 0$), demography has little effect on the power of the association test or the frequencies of the causal variants (**Fig. S2.2B**, **Fig. S2.2D**, and **Fig. S2.2F**).

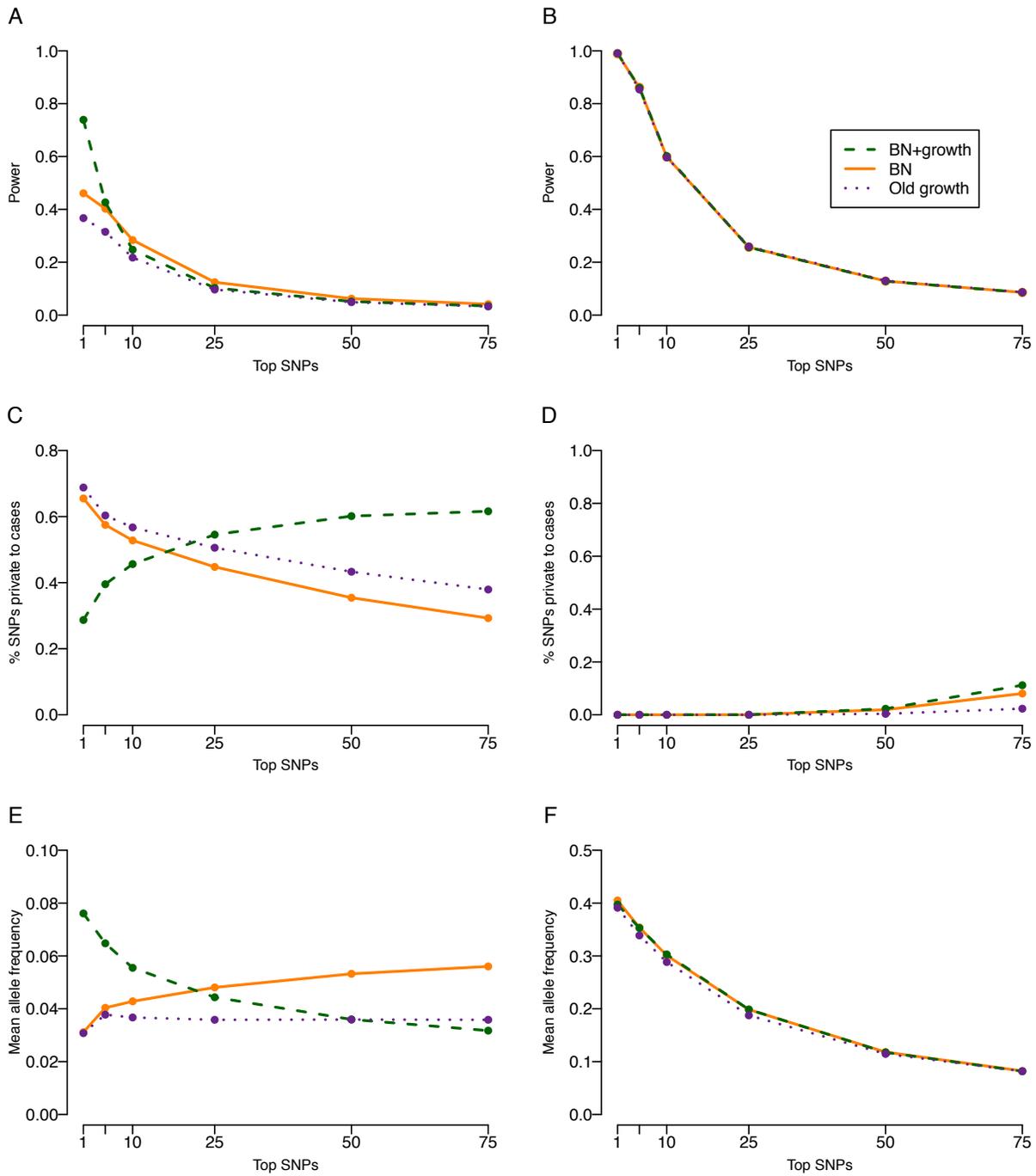

**Figure S2.2: Properties of single-marker association tests for causal variants in a study of 1000 cases and 1000 controls for the SNPs that explain the most $V_A$ per simulation replicate (*x*-axis).** Statistics were calculated for each simulation replicate and then averaged over the 1000 simulation replicates. **(A, C, E)** A SNP's effect on the trait is correlated with its effect on fitness ($\tau = 0.5$). **(B, D, F)** A SNP's effect on the trait is independent of its effect on fitness ($\tau = 0$). **(A-B)** Power. **(C-D)** Proportion of causal variants that were segregating in cases only. **(E-F)** Average allele frequency of the causal variants.

*Power as a function of allele frequency*

Next I examine the power of the single-marker association tests as a function of the allele frequency of the SNP (**Fig. S2.3**). When a mutation's effect on fitness is correlated with its effect on the trait ($\tau = 0.5$), I find that power is highest for SNPs with allele frequencies between 5-10% (**Fig. S2.3A**). Additionally, power is slightly higher in the population that did not expand as compared to that in the expanded population. Power is low for rare SNPs, as expected. Power is also low for very common SNPs (>10%) because, under this model, these SNPs also have the smallest effect sizes, which leads to a decrease in power. While recent population growth only has a subtle effect on the power of the association test when conditioning on allele frequency, it has a substantial effect on the number of rare causal variants. As expected, population growth increases the number of rare causal variants as compared to a population that did not expand (**Fig. S2.3C**). These are the variants that are very difficult to detect via single-marker association tests. Similar to previous results, when a mutation's effect on fitness is not correlated with its effect on the trait ($\tau = 0$), I find that demography has little effect on the power of the association test, and that the power of the test is highest for common variants (allele frequency >10%; **Fig. S2.3B**). Under this model, effect sizes are not correlated with allele frequencies (**Fig. S2.3B**), and as such, common variants are just as likely to have large effects as are rare variants. Thus, power is highest to detect common variants. Again, however, recent population growth increases the number of rare causal variants in the population (**Fig. S2.3D**), which will be difficult to detect using single marker association tests.

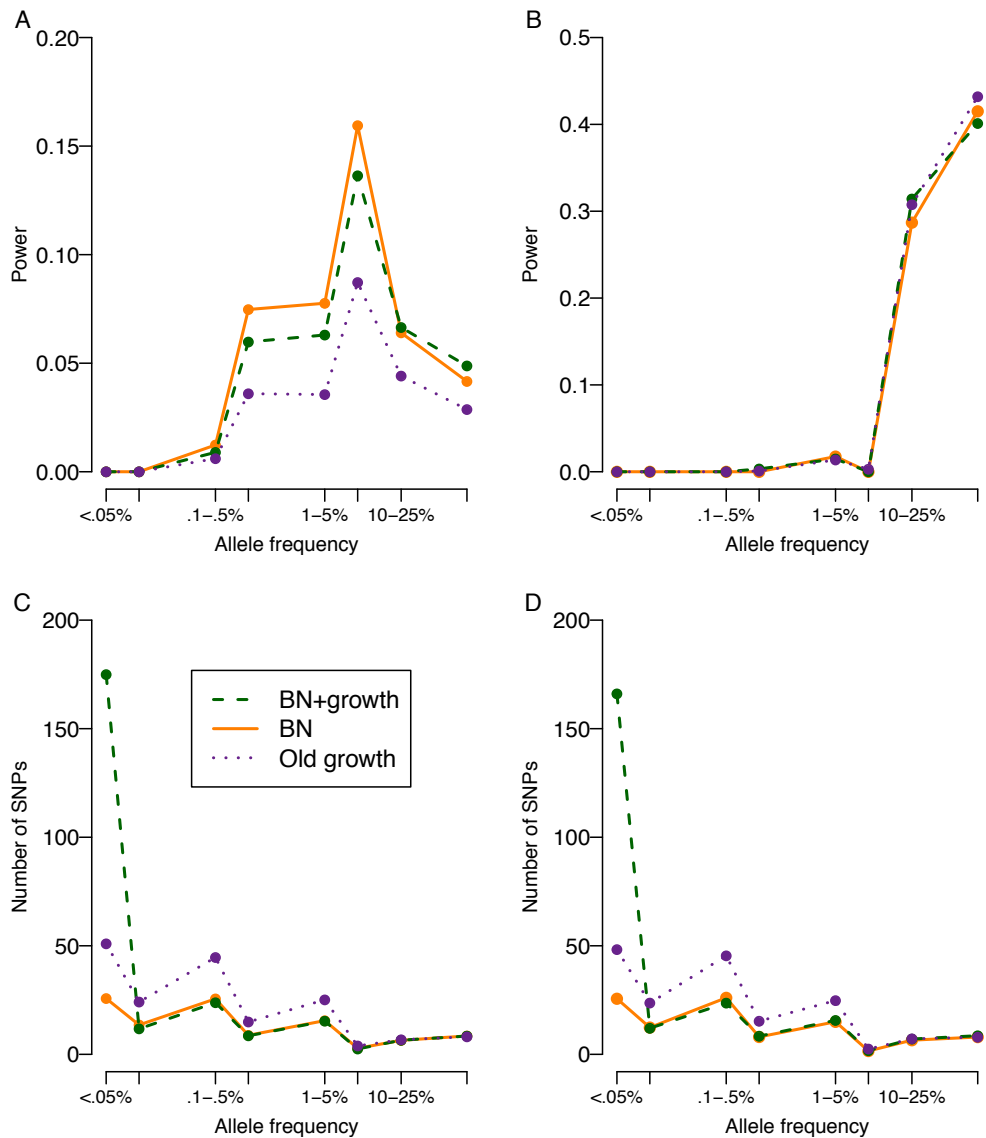

**Figure S2.3: Properties of single-marker association tests for causal variants in a study of 1000 cases and 1000 controls, binned by the causal variant allele frequency (*x*-axis).** Statistics were calculated for each simulation replicate, and then were averaged over the 1000 simulation replicates. **(A, C)** A SNP's effect on the trait is correlated with its effect on fitness ($\tau = 0.5$). **(B, D)** A SNP's effect on the trait is independent of its effect on fitness ($\tau = 0$). **(A, B)** Power. **(C, D)** Number of casual variants at different allele frequencies. Note that recent growth (BN+growth) drastically increases the number of rare causal variants (dashed green lines).

*Power as a function of odds ratio (OR)*

I also examine the power of the association test as a function of the estimated odds ratio (OR) computed from the case-control study (**Fig. S2.4**). When a mutation's effect on fitness is correlated with its effect on the trait ($\tau = 0.5$), power is highest for those SNPs with an OR close to 10 (**Fig. S2.4A**). Power decreases as the effect sizes decrease, and there is essentially no difference in power across the different models of population history. Many SNPs were present only in cases. An OR calculated for such SNPs would be infinite. However, power is low to detect such variants because they are typically at low frequency, and single-marker tests are underpowered to detect such variants [1]. Recent population growth increases the number of such mutations (**Fig. S2.4C**). When a mutation's effect on fitness is independent of its effect on the trait ($\tau = 0$), power is highest for SNPs with ORs between 1.5 and 2. SNPs with higher ORs are typically at low frequency in the population, reducing the power to detect them (**Fig. S2.4B**). Though the effect size on the liability scale in the population is not correlated with allele frequency (**Fig. S10B**), low-frequency SNPs tend to have larger ORs, simply because they are more likely to show larger relative differences in frequency between cases and controls (**Fig. S10D**). Again, population history has little effect on power (**Fig. S2.4B**). Recent growth also increases the number of mutations that are present only in cases and that appear to have ORs of infinity (**Fig. S2.4D**). Growth would also increase the number of mutations present only in controls that would have ORs of 0. However, the median OR is still >1, reflecting the fact that cases carry more variants that are not carried by controls (rather than vice versa). This is expected as mutations were expected to increase risk of disease and as such, cases are expected to carry more of them.

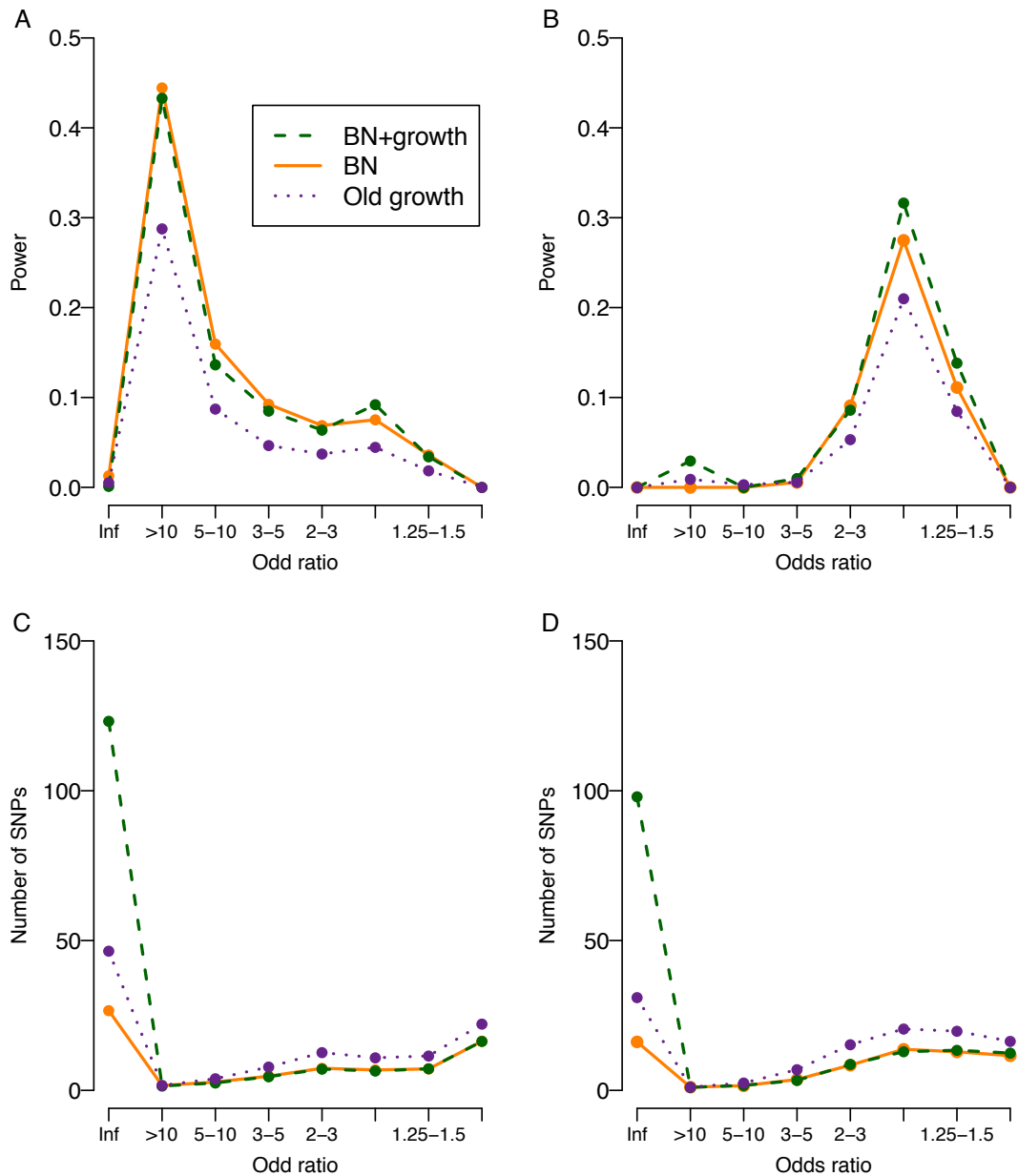

**Figure S2.4: Properties of single-marker association tests for causal variants in a study of 1000 cases and 1000 controls, binned by the estimated odds ratio (*x*-axis).** Statistics were calculated for each simulation replicate and were averaged over the 1000 simulation replicates. An odds ratio of "Inf" refers to those variants that were found only in cases. **(A, C)** A SNP's effect on the trait is correlated with its effect on fitness ($\tau = 0.5$). **(B, D)** A SNP's effect on the trait is independent of its effect on fitness ($\tau = 0$). **(A, B)** Power. **(C, D)** Number of casual variants at different allele frequencies. Note that recent growth drastically increases the number of causal variants found exclusively in cases (green lines; BN+growth).

# REFERENCES

1. Kryukov GV, Shpunt A, Stamatoyannopoulos JA, Sunyaev SR. (2009) Power of deep, all-exon resequencing for discovery of human trait genes. Proc Natl Acad Sci U S A 106: 3871-3876.